\documentclass{aa}  
\usepackage{booktabs}
\usepackage{longtable}
\usepackage[colorlinks=true, linkcolor=blue, citecolor=blue, urlcolor=blue]{hyperref}
\usepackage{siunitx} 
\usepackage{todonotes}
\usepackage{txfonts}
\usepackage{titlesec}
\newcommand{\xmm}{{\it XMM-Newton}\xspace}
\newcommand{\gaia}{{\it Gaia}\xspace}

\begin{document} 
\twocolumn

   \title{Unveiling the soft X-ray source population towards the inner Galactic disk with \textit{XMM-Newton}}

   \author{Tong Bao
          \inst{1},
          Gabriele Ponti\inst{1,2,3},
		  Frank Haberl \inst{2},
		  Samaresh Mondal \inst{4}
		  \and 
		  Mark R. Morris \inst{5}
		  \and
		  Kaya Mori\inst{6}
		  \and
		  Shifra Mandel\inst{6}
		  \and
		  Xiao-jie Xu\inst{7}
          }

   \institute{$^1$ INAF -- Osservatorio Astronomico di Brera, Via E. Bianchi 46, 23807 Merate, Italy
              \\\email{tong.bao@inaf.it}
              \\
         $^2$
             Max-Planck-Institut für Extraterrestriche Physik, Gießenbachstraße 1, 85748 Garching, Germany\\
         $^3$
          Como Lake Center for Astrophysics (CLAP), DiSAT, Università degli Studi dell’Insubria, via Valleggio 11, I-22100 Como, Italy\\
         $^4$ Department of Astronomy, University of Illinois, 1002 W. Green St., Urbana, IL 61801, USA\\
         $^5$ Department of Physics and Astronomy, University of California, Los Angeles, CA, 90095-1547, USA\\
         $^6$ Columbia Astrophysics Laboratory, Columbia University, New York, NY 10027, USA\\
         $^7$
          School of Astronomy and Space Science, Nanjing University, Nanjing 210046, China
             }

   \date{Received August 4, 2025; accepted October 26, 2025}
   \titlerunning{Soft X-ray source towards the inner Galactic disk}
   \authorrunning{Bao et al.}
 
  \abstract
   {Across the Galactic disk lies a diverse population of X-ray sources, including coronally active stars and accreting compact objects. While high-luminosity sources are well characterized, the fainter end of the population remains poorly understood due to sensitivity limitations in previous X-ray surveys.}
   {We aim to classify and characterize faint X-ray sources detected in the eROSITA All-Sky Survey (eRASS1) towards the inner Galactic disk using deeper \xmm observations. By combining X-ray spectral analysis with Gaia counterparts, we assess the representativeness of the eRASS1 catalog for the broader Galactic X-ray population.}
  {We analyzed 189 eRASS1 X-ray sources towards the inner Galactic disk, using deep {\it XMM-Newton} observations (typical exposure of 20~ks) covering the region $350\degr < l < 360\degr$ and $-1\degr < b < 1\degr$.
  Source classification was carried out by combining X-ray spectral fitting in the 0.2--10~keV band with Gaia astrometric and photometric data.}
  {Approximately 74\% of X-ray sources detected by eROSITA towards the inner Galactic disk are coronal sources (primarily active stars and binaries), while 8\% are wind-powered massive stars and 18\% are accreting compact objects. We propose an empirical hardness-ratio cut ($\mathrm{HR} > -0.2$, with HR defined using the 0.5–2 and 2–8 keV bands) to efficiently identify non-coronal sources within the eRASS1 catalog. By stacking classified sources and comparing with the Galactic ridge X-ray emission (GRXE), we estimate that $\sim$~6\% of the GRXE flux can be resolved into point sources in the 0.5--2.0 keV band above the eRASS1 flux limit of $\sim$~5$\times$~10$^{-14}$~erg~cm$^{-2}$~s$^{-1}$, with soft-band emission dominated by active stars and hard-band flux primarily from X-ray binaries.}
    {Our results demonstrate that the eRASS1 catalog towards the inner Galactic disk is dominated by coronal sources but retains a non-negligible population of massive stars and accreting compact objects that can be effectively identified with X-ray color selection.}

   \keywords{X-rays: stars -- X-rays: binaries -- novae, cataclysmic variables -- Galaxy: disk -- Galaxy: stellar content}

   \maketitle

\section{Introduction}

The study of Galactic X-ray sources has advanced significantly since the early all-sky surveys conducted by missions such as \textit{HEAO-1} and \textit{ROSAT}, which revealed that the most luminous X-ray sources in the Milky Way are primarily X-ray binaries (XRBs) and supernova remnants \citep{1986ApJ...311..258A,Truemper1993}. These objects dominate the high-flux end of the X-ray sky, and their nature has been well characterized through follow-up multi-wavelength identification programs.

While the high-luminosity end of the X-ray source population is now relatively well understood, many different classes of fainter sources contribute to the Galactic X-ray landscape \citep{Sazonov2006}. The population of low- to intermediate-luminosity X-ray sources (typically defined as $L_X \lesssim 10^{34}~\mathrm{erg~s^{-1}}$) remains far less explored, with our current knowledge largely limited to nearby representatives in the solar neighborhood \citep{Pretorius2012,Pretorius2013,Pala2020}. 
These fainter sources, comprising cataclysmic variables (CVs), active binaries (ABs) and even some quiescent X-ray binaries, are believed to represent the dominant contributors to the so-called Galactic Ridge X-ray Emission (GRXE), a diffuse X-ray emission concentrated toward the Galactic plane \citep{Worrall1983, Revnivtsev2006, Revnivtsev2009, Hong2012, Reis2013, Muno2004}.
The Galactic disk, which harbors a complex mixture of stellar populations, star-forming regions, and dense interstellar medium, presents an ideal laboratory for studying these faint X-ray emitters \citep{Degenaar2012,ponti2015, Hong2016,Wang2021}.

The early \textit{ROSAT} all-sky maps reveal a broad soft X-ray distribution (1.5 keV) across the Galactic disk. 
After that, \textit{ASCA} observations of the Galactic plane detected $\sim$0.8~keV ionized plasma in the GRXE spectra \citep{Kaneda1997,Tanaka2002} and the harder component ($kT \sim 7$~keV) is generally associated with accreting binaries including cataclysmic variables, and active binaries, as revealed by subsequent X-ray surveys \citep{Muno2004, Revnivtsev2009, Morihana2013, Yamamoto2023}.

Deep \textit{Chandra} observations of the Galactic center, bulge, and plane have resolved a significant fraction of the broad-band (2–10keV) GRXE into discrete sources, particularly the Fe line emission at $\sim$6–7 keV \citep{Revnivtsev2009}. In contrast, the origin of the softer thermal emission (below $\sim$2~keV) remains less well constrained. As suggested by early XMM Galactic Plane Survey results \citep{Hands2004}, it is likely dominated by nearby, low-luminosity sources, whose contributions have not yet been systematically quantified.

The advent of the eROSITA mission has revolutionized X-ray astronomy with its unprecedented all-sky coverage and improved sensitivity \citep{Merloni2012,Sunyaev2021}. The eROSITA All-Sky Survey (eRASS; \citealp{eRASS1}) has detected a large number of X-ray sources in the Galactic plane, significantly increasing the known population. However, despite its wide coverage, eROSITA's relatively shallow depth and limited spatial resolution in dense Galactic fields can lead to source confusion and difficulties in obtaining detailed spectral information for fainter detections. This often leaves the precise nature of many eRASS1 sources in these complex regions ambiguous, hindering a complete census and characterization of the faint X-ray population.

Our ongoing \emph{XMM-Newton} Heritage Survey of the Galactic Plane, however, was designed to uniformly cover the Galactic plane region ($350\degr < l < 7\degr$, $|b| < 1\degr$) with $\sim 20$~ks exposures, achieving an order-of-magnitude improvement in sensitivity compared to eROSITA.
The deeper exposures provided by \xmm allow the detection and characterization of fainter sources below the eRASS1 detection limit and, crucially, enable the extraction of high-quality X-ray spectra for sources detected by both missions. These detailed spectra provide vital diagnostic information (e.g. plasma temperature, absorption, presence of emission lines) that is often unavailable from eROSITA data alone. 
In this work, we focus on characterizing the eRASS1 flux-limited sample in combination with deep {\it XMM-Newton} observations and detailed multi-wavelength classifications. This allows us to dissect the observed soft-band GRXE into contributions from distinct source populations.

Here we present a comprehensive study of the soft X-ray source population towards the inner Galactic disk, focusing on the sources detected by both eROSITA and \xmm. By combining X-ray spectral analysis, \gaia DR3 astrometry and photometry, we aim to (1) classify the nature of faint Galactic X-ray sources; (2) validate empirical classification schemes based on eROSITA observables; and (3) quantify the contribution of resolved stellar sources to the observed GRXE.
The paper is organized as follows: Section~\ref{sec:catalog} describes the construction of our X-ray source catalog and the XMM/eROSITA cross-matching. 
Section~\ref{sec:gaia} details the identification of Gaia counterparts and the reliability of classification done by the \textit{HamStar} method \citep{hamstar}. 
In Sect.~\ref{sec:spec}, we outline the X-ray data reduction and spectral analysis. 
Source classification based on multi-wavelength diagnostics is presented in Sect.~\ref{sec:class}.
In Sect.~\ref{sec:discuss}, we propose a simple empirical hardness ratio cut to select accreting compact objects from the eROSITA Galactic plane catalog, and assess the contributions of various resolved source classes to the GRXE spectra.

\section{XMM-Newton observations and source catalog}
\label{sec:catalog}
We are in the process of completing the {\it XMM-Newton} Heritage survey towards the inner Galactic disk, which covers the disk region in the range of $350\degr < l < 7\degr$ and $ -1\degr < b < 1\degr$. The average exposure is approximately 20 ks per tile, providing a factor of $\sim$ 10 improvement in sensitivity over eRASS1.

\subsection{4XMM DR14s}
We utilized the XMM-Newton Serendipitous Source Catalog from Stacked Observations (4XMM-DR14s; \citealp{4XMMDR14s}). This catalog was constructed from simultaneous source detection on overlapping observations between 2001 and 2024, except for a 15\arcmin\ region about the Galactic Center, which was excluded from the processing for technical reasons. 
Of the 606 observations from our ongoing XMM-Newton Heritage Survey, 562 are included in this study. The remaining observations were either conducted after 2024 or lie within the excluded Galactic Center region. We applied rigorous quality cuts as follows:  
\begin{itemize}
  \item STACK\_FLAG $\leq$ 1 (minimizes stacked detection artifacts)  
  \item EP\_DET\_ML $\geq$ 6 (false-positive probability $<0.1\%$)  
  \item EXTENT = 0 (include point sources only)  
\end{itemize}
The final sample comprises 15655 sources falling within the footprint of our {\it XMM-Newton} Heritage survey, where 9533 of them are observed by more than one observation. 
The observation data files were processed using the \xmm Science Analysis System (SAS, v22.1.0\footnote{https://www.cosmos.esa.int/web/xmm-newton/sas-release-notes-2210}). We used the task \emph{evselect} to construct a high-energy background light curve (energy between 10 and 12 keV for EPIC-pn \citep{struder2001} and above 10 keV for EPIC-MOS1/MOS2 \citep{turner2001}) by selecting only PATTERN==0. The background light curve was used to filter high-background flaring activity and to thereby identify good time intervals. 

\subsection{eRASS1 catalog}
The eROSITA telescope array aboard the \textit{Spektrum Roentgen Gamma (SRG)} satellite began surveying the sky in December 2019 \citep{Predehl2021,Sunyaev2021}.
Here we utilized the eRASS1 catalog \citep{eRASS1}, which contains sources located in the western Galactic hemisphere, using the data acquired in the first six months of survey operations (eRASS1; completed June 2020).
We selected sources from the eRASS1 main catalog, which includes sources detected in the 0.2–2.3 keV band, with DET\_LIKE\_0 $\geq$ 10 and EXT\_LIKE > 0 to reduce spurious detections lower than 1\% and remove extended sources. 
In total, there are 182 sources left within the footprint of our {\it XMM-Newton} Heritage survey. 

We also examined the eRASS1 Hard catalog (2.3--5 keV band) and identified 14 hard-band sources located within the sky coverage of our study. Among them, seven have ``strong'' associations, as defined in \citet{eRASS1}, and correspond to bright sources already included in our selected sample from the Main catalog. The remaining seven exhibit only ``weak'' or no associations. 
These are labeled as ``Hard-only'' sources in the lower section of Table~\ref{tab:specsrc}.
In total, 189 eRASS1 sources fall within the footprint of our {\it XMM-Newton} Heritage survey.

\subsection{Cross-matching between eRASS1 and XMM sources}
\begin{figure}
  \centering
   \includegraphics[width=\hsize]{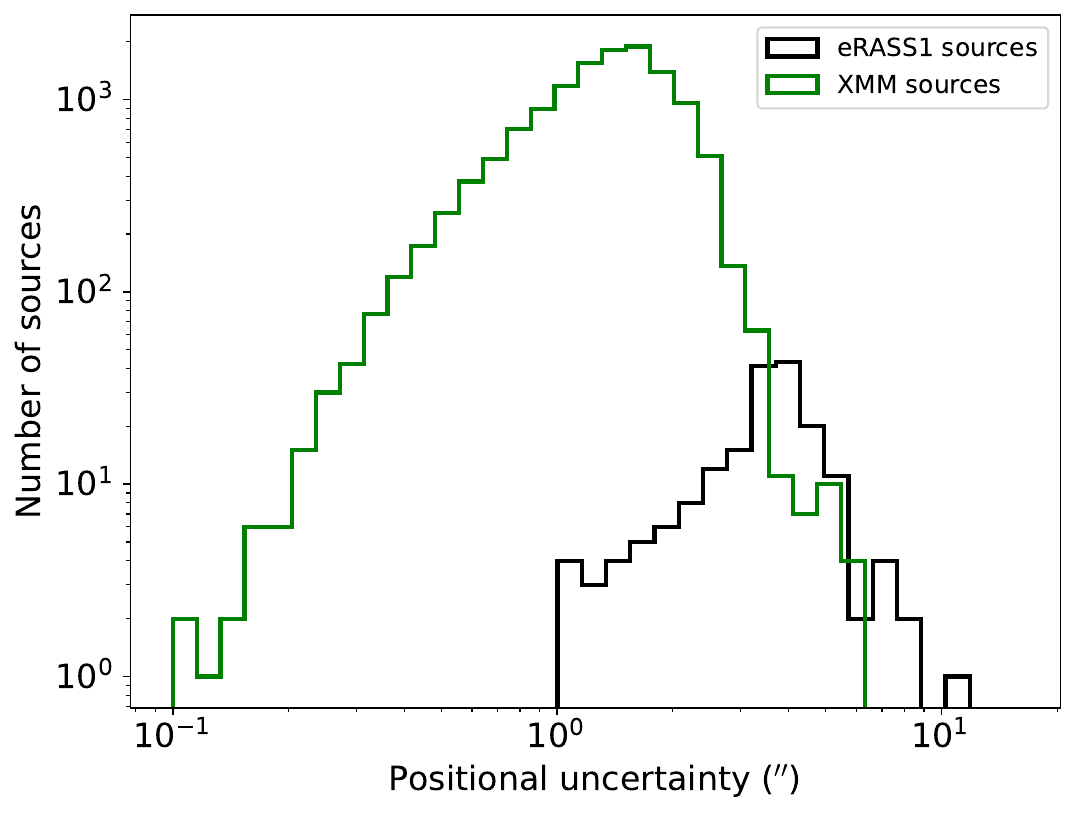}
      \caption{Positional uncertainty comparison of eRASS1 and XMM-Newton sources. This histogram shows the 1$\sigma$ positional uncertainties for sources in the eRASS1 (black) and \xmm (green) catalogs.}
       \label{fig:POSerr}
\end{figure}

The positional uncertainty of X-ray sources is a crucial parameter for associating them with multi-wavelength counterparts. 
We present the distribution of 1$\sigma$ positional uncertainties for the selected 189 eRASS1 point sources.
For comparison, we also show the 1$\sigma$ positional uncertainties for the selected XMM sources from the 4XMM-DR14s catalog \citep{4XMMDR14s}.
The mean positional uncertainties for eRASS1 and XMM sources are $\sim$ 3.63\arcsec and $\sim$ 1.39\arcsec, respectively, as shown in Fig.~\ref{fig:POSerr}.

\begin{figure}
  \centering
   \includegraphics[width=\hsize]{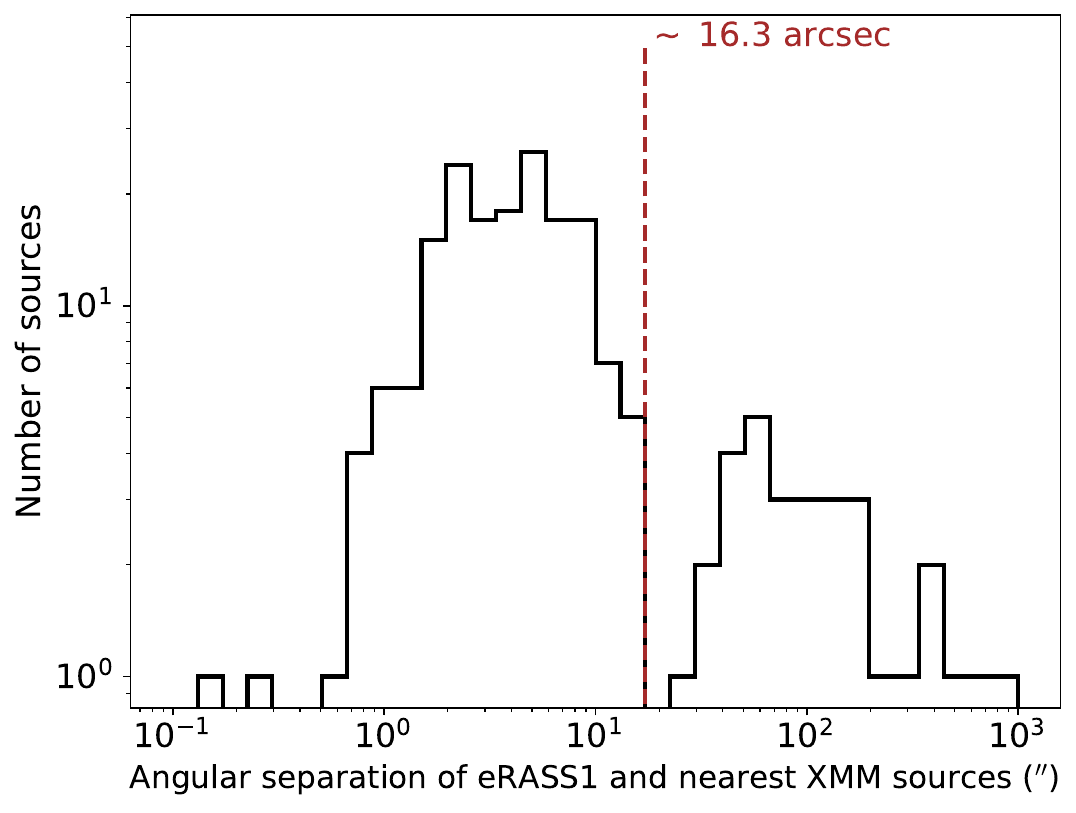}
      \caption{Angular separation between eRASS1 and nearest XMM-Newton sources.
      The prominent bimodal distribution indicates a natural separation threshold at approximately $\sim$16.3\arcsec, used to distinguish true associations from random alignments.}
       \label{fig:SEPhist}
\end{figure}

We further analyze the angular separation between the eRASS1 sources and their nearest XMM counterparts. The resulting distribution shows a distinct bimodal structure with a break at around $\sim$ 16.3\arcsec, corresponding to approximately 4.2 $\sigma$ of the combined positional uncertainties of the two catalogs. This separation threshold was adopted as a conservative criterion to ensure the inclusion of all likely true counterparts.  
As shown in Fig.~\ref{fig:SEPhist}, the bimodal distribution clearly distinguishes matched pairs from chance coincidences. Applying this separation threshold, we find that 158 of the 189 eRASS1 sources have counterparts in the XMM catalog, while the remaining 31 lack matches.
Although the largest separations of our XMM-eRASS1 matches reach over 16.3\arcsec, we should note that only 5\% (8 of 158) pairs have separation over 11.3\arcsec, approximately 3$\sigma$ of the combined mean positional uncertainties. For those eight sources, their positional errors are also larger than the mean positional uncertainties due to their faintness.
Therefore, we include all 158 matched pairs in our selected sample.

\begin{figure}
  \centering
   \includegraphics[width=\hsize]{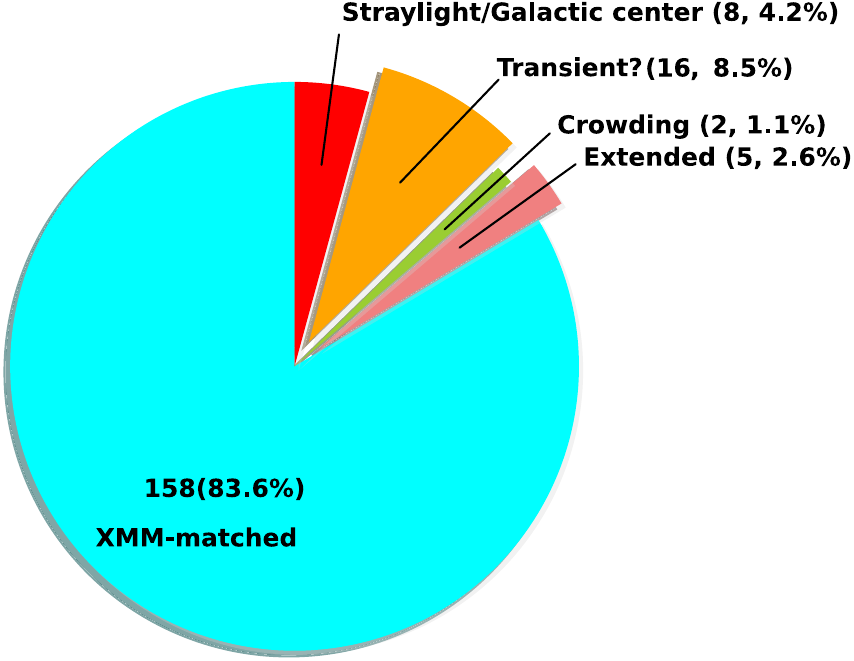}
      \caption{Cross-matching results between eRASS1 and XMM sources. Among 189 eRASS1 sources within the \xmm footprint, 158 have reliable XMM counterparts. The remaining 31 are categorized as potentially extended sources, source confusion in crowded regions, transient candidates, or straylight contamination/Galactic center sources and thus are outside the XMM filtered catalog.}
       \label{fig:eXMMpie}
\end{figure}

As illustrated in Fig. \ref{fig:eXMMpie}, the 31 unmatched sources can be categorized as follows: Five of them appear to be extended sources that were misidentified as point sources; two of them lie in crowded regions (and hence may not be detected due to possible source confusion -- i.e. multiple X-ray sources are detected as a single source by eROSITA); 
eight of them are located near the Galactic center or areas affected by stray light; and the remaining 16 are candidate transient sources. 
Hereafter we focus on the 158 sources with both XMM and eROSITA detections.

\section{Gaia counterparts of eRASS1 sources}
\label{sec:gaia}

To enable a more robust classification of our X-ray sources, we cross‐matched the 158 eRASS1 sources with optical counterparts from \gaia\,DR3 \citep{Gaiamission,GaiaDR3} using their \xmm positions, and compared the results to those using the original eROSITA coordinates.

Assuming that the positional uncertainties of both X-ray and optical measurements follow Gaussian distributions centered on the true source location, the angular separation: \( r \), between X-ray and optical positions is expected to follow a Rayleigh distribution:
\begin{equation}
p\left(r ; \sigma_{\mathrm{sep}}\right)=\frac{r}{\sigma_{\mathrm{sep}}^2} \exp \left(-\frac{1}{2} \frac{r^2}{\sigma_{\mathrm{sep}}^2}\right)
\end{equation}  
where \( \sigma_{\mathrm{sep}} \) is the combined positional uncertainty.
Fitting the histograms of eROSITA–\gaia and XMM–\gaia separations yields \( \sigma_{\mathrm{sep}} = 3.44'' \) between \gaia and eRASS1, and \( \sigma_{\mathrm{sep}} = 0.89'' \) between \gaia and \xmm, reflecting the improved astrometric accuracy of the latter due to better angular resolution and longer exposure, as shown in Fig.~\ref{fig:eXMMGAIA_sep}.
Out of 158 sources, we find that 146 have \gaia counterparts within \(4''\), a conservative threshold of about 4.5 $\sigma_{\mathrm{sep}}$ we applied here\footnote{Most of the matched pairs (95\%, 139 of 146) have separations within 3$\sigma$ ($\sim$ 2.67\arcsec) of the mean separation between \gaia and \xmm.}.
There are 12 sources, including three “hard‐only” objects, with no Gaia counterpart within 4\arcsec.

These sources may be undetected by \gaia due to high X-ray–to–optical flux ratios, as seen in X-ray binaries or AGN, or because they are distant, heavily obscured, or other intrinsically faint conditions that place them below the \gaia sensitivity limits.

\subsection{Comparison to the \textit{HamStar} catalog}
\label{sec:hamstar}
\citet{hamstar} introduced the \textit{HamStar} method, a Bayesian framework developed to identify coronal X-ray sources in the eRASS1 catalog by associating them with \gaia DR3 counterparts. This method estimates the probability of a source being coronal by incorporating both geometric parameters, such as angular separation and positional uncertainty, and source properties, including X-ray flux, optical color, and distance. 

\begin{figure}
  \centering
   \includegraphics[width=1.04\hsize]{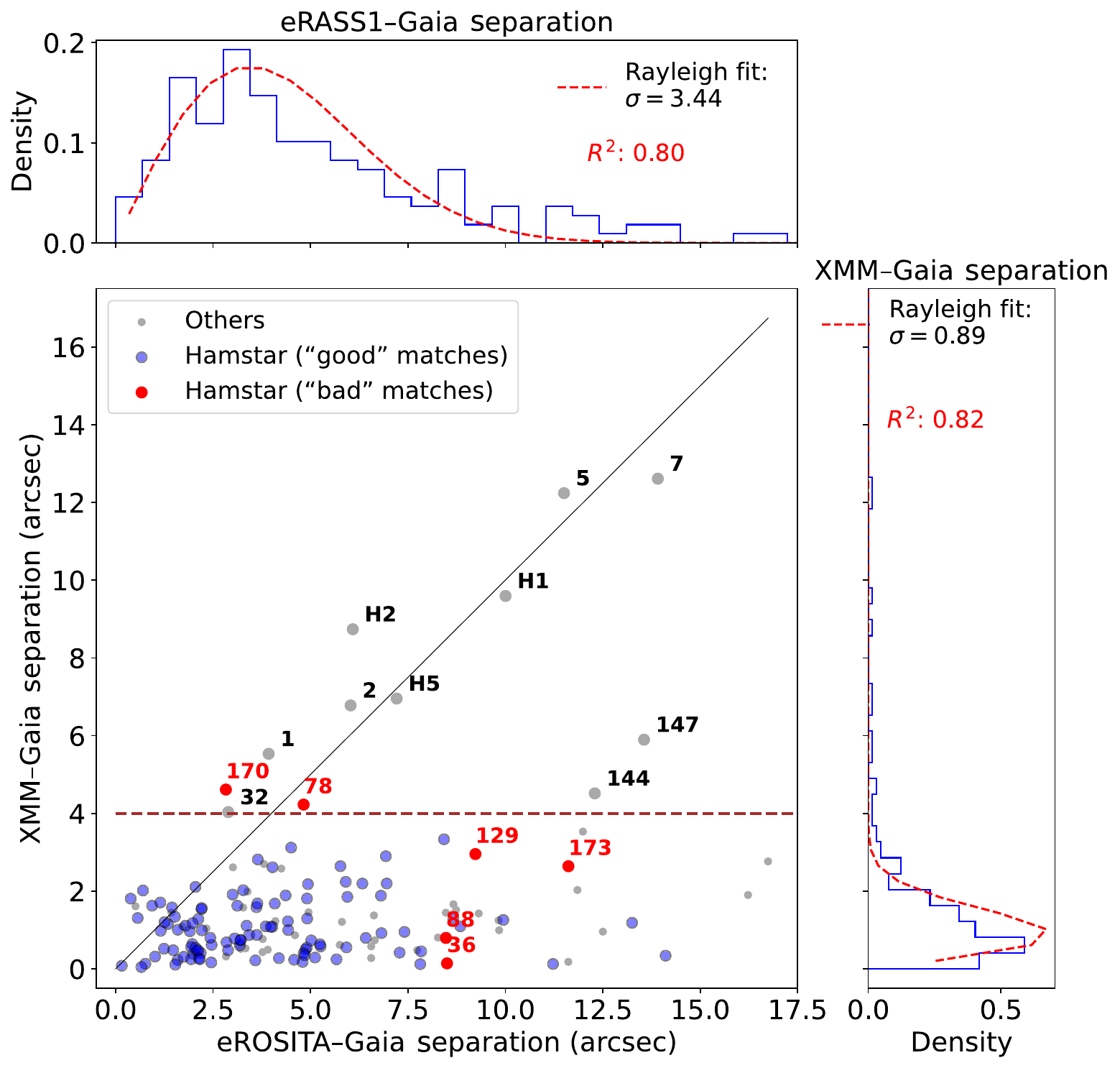}
      \caption[width=\hsize]{
      Angular separation distributions for eRASS1, \xmm, and \gaia matched sources. Histograms (top and right panels) display eRASS1-\gaia and XMM-\gaia separations, respectively. Rayleigh fits (red dashed lines) yield characteristic uncertainties of $\sigma_{\rm sep}$ = 3.44\arcsec\ for eRASS1-\gaia pairs and $\sigma_{\rm sep}$ = 0.89\arcsec\ for XMM-\gaia pairs.    
      Main panel: eRASS1-\gaia versus XMM-\gaia separation with the y=x equality shown as a black dashed line. ``Good'' Hamstar matches are blue dots. ``Bad'' Hamstar matches are red dots with their source indices. Sources without a Hamstar match are represented by dark gray dots and indices of vertical outliers (XMM-\gaia separations $>$ 4\arcsec) are labeled in black.}
       \label{fig:eXMMGAIA_sep}
\end{figure}

Within our sample of 158 sources, 108 were previously classified as coronal by the \textit{HamStar} algorithm based on their eROSITA positions. 
After re-performing the cross-match using \xmm coordinates, we find that 102 of these sources have the same \gaia counterpart as originally identified by \textit{HamStar}, with the X-ray–to–optical separations significantly reduced to within 4\arcsec. We refer to these as ``good'' matches, indicated by the blue dots in the main panel of Fig.~\ref{fig:eXMMGAIA_sep}.

The remaining six \textit{HamStar} sources are classified as ``bad'' matches. Unlike the ``good'' matches, whose X-ray–optical separations typically decreased when using the more accurate \xmm positions, these sources show the opposite behavior: their previously assigned \gaia counterparts from the \textit{HamStar} catalog exhibit larger separations after applying the \xmm astrometry. Therefore, we consider these to be likely false associations.
Among them, four are now more closely matched to different \gaia counterparts with separations less than 4\arcsec\ from the original eROSITA-Gaia matches, classified as ``Different match for Hamstar'' in Fig.~\ref{fig:eXMMpie}. The remaining two (Src~78 and Src~170) have no Gaia match within 4\arcsec. These are shown as red dots in Fig.~\ref{fig:eXMMGAIA_sep}. 

\begin{figure}
  \centering
   \includegraphics[width=\hsize]{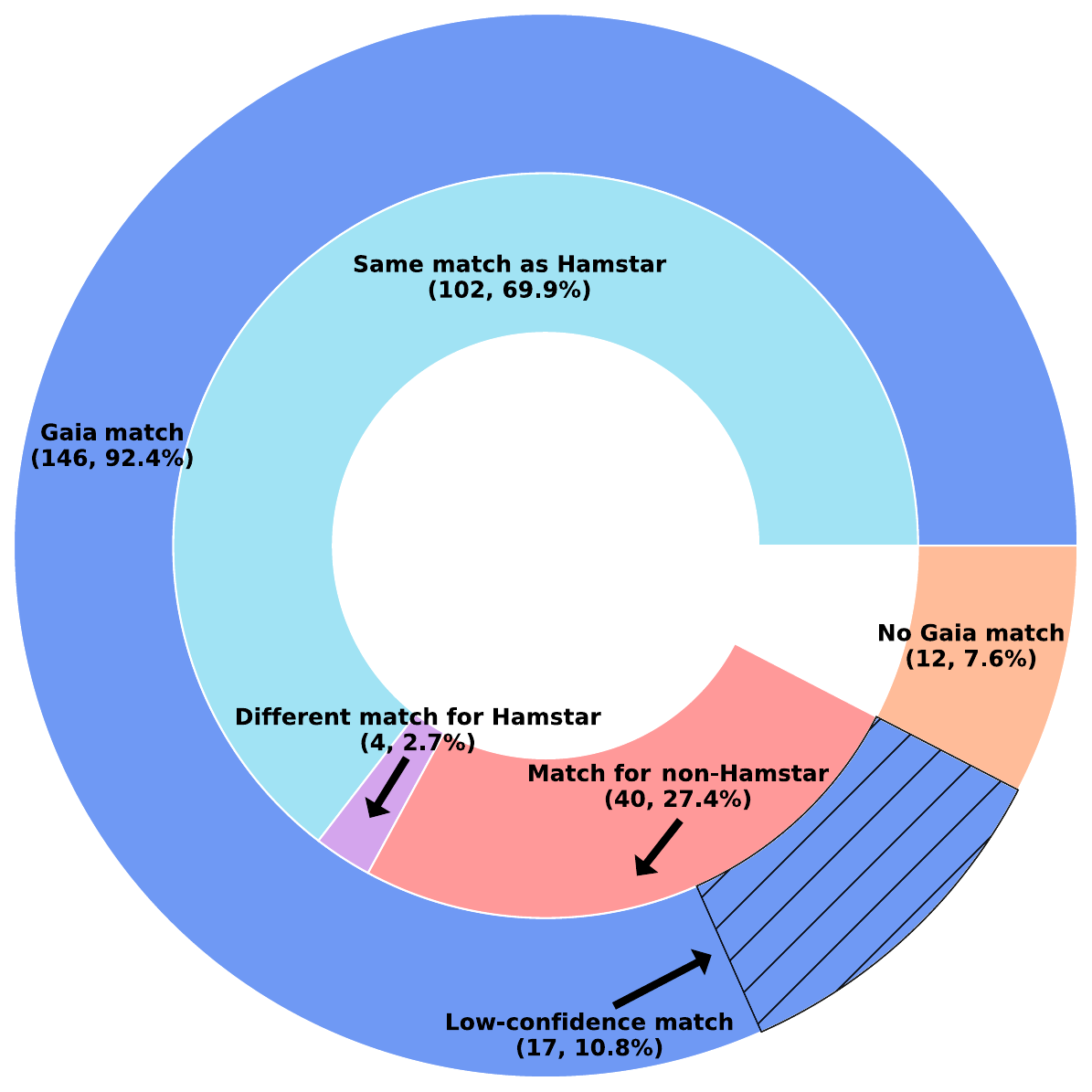}
      \caption[width=\hsize]{\gaia matching results for the 158 \xmm-detected eRASS1 sources.}
       \label{fig:xmmgaia}
\end{figure}

The remaining 50 sources were not included in the \textit{HamStar} catalog primarily because the method incorporates not only positional coincidence but also multi-wavelength criteria, including X-ray flux, optical color, and parallax within a Bayesian framework. 
Consequently, a \gaia source within the matching radius may be rejected if its properties (e.g., X-ray–to–optical flux ratio, color) deviate from those typical of coronal emitters. This conservative approach favors reliability but may miss valid associations.
By re-matching with improved XMM positions, we recover additional plausible \gaia counterparts (40 out of 50) that were previously excluded due to stricter priors of \textit{HamStar}. The enhanced X-ray astrometry of \xmm leads to more precise associations, reducing the number of ambiguous or false matches. The overall \gaia matching results are summarized in the pie chart in Fig.~\ref{fig:xmmgaia}.

\subsection{False-match probability}
\label{sec:falsematch}

Due to significant spatial variation in the stellar density across our field of view, the likelihood of falsely associating an X-ray source with a random \gaia counterpart can vary substantially. To account for this, we estimated the probability of a chance (i.e., false) match for each \xmm source based on the local density of \gaia detections.

For each X-ray source, we calculated the surface density of \gaia sources within a 1\arcmin\ radius, denoted as $\rho_{Gaia}$ (in arcsec$^{-2}$). Assuming a uniform background distribution, the probability that at least one unrelated \gaia source lies within a circular matching radius $r_{\rm match}$ is given by the Poisson expression \citep{1992MNRAS.259..413S}:
\begin{equation}
P_{\rm false} = 1 - \exp\left( -\pi \rho_{Gaia} \cdot r_{\rm match}^2 \right).
\end{equation}

This formula provides a statistical estimate of the likelihood that a given \gaia match is spurious, based purely on positional coincidence and local source density. We computed $P_{\rm false}$ for all 146 \gaia-matched sources in our sample and listed their results in Table~\ref{tab:specsrc}, providing a quantitative measure of counterpart reliability.
The majority of sources ($\sim$ 88\%, 129 out of 146) have $P_{\rm false} < 0.05$, suggesting high-confidence associations. The remaining 17 sources (12\%), primarily located in crowded or high-density regions, exhibit higher false-match probabilities. These ``Low-confidence matches'' are discussed in detail in Sect. \ref{sec:class}.

While we adopt $P_{\rm false} < 0.05$ as a threshold to assess the reliability of \gaia counterparts, we choose to retain all associations within 4\arcsec\ in our analysis, regardless of their individual $P_{\rm false}$ values. 
This approach is intended to preserve sample completeness and maintain internal consistency across the dataset, particularly when comparing sources with and without prior classifications.
Moreover, many of the sources with higher $P_{\rm false}$ values are located in crowded Galactic regions, where source confusion is inherently high and positional offsets alone may not fully account for the counterpart uncertainty. 
By including all matches within 4\arcsec, while explicitly flagging those with $P_{\rm false} > 0.05$, we ensure that potentially genuine associations are not prematurely excluded, while allowing users to apply more stringent reliability criteria based on specific scientific objectives. 

Overall, 146 of the 158 eRASS1 sources detected with \xmm have \gaia counterparts within 4\arcsec, including 102 that are consistent with \textit{HamStar} classifications. The remaining 44 \gaia associations, made possible by the improved astrometry from XMM, supplement the \textit{HamStar} catalog and enhance the completeness of our sample.
Adopting a conservative threshold of $P_{\rm false} < 0.05$ as a criterion for secure identification, we estimate the reliability of the \gaia-matched counterparts in our sample to be approximately 88\%.

\section{Spectral analysis}
\label{sec:spec}
The nature of most eRASS1 sources remains uncertain due to the limited photon counts, which hinder reliable spectral characterization. 
While the \textit{HamStar} method offers a promising approach by applying photometric measurements of potential optical counterparts to identify coronal sources, the reliability of this classification has yet to be systematically validated. 
Our deep \xmm survey towards the inner Galactic disk, covering a subset of eRASS1 sources, enables not only improved X-ray positions but also detailed X-ray spectral analysis, thereby allowing for more robust source classification.

For each source, we extracted the source and background spectra using the SAS task \emph{evselect}. 
Events with PATTERN $\leq$ 4 for EPIC-pn and PATTERN $\leq$ 12 for MOS1/MOS2 detectors were selected. 
The source products were extracted from a circular region with a radius of 25\arcsec\ centered on the source position, while the background was extracted from an annular region centered on the same position, with inner and outer radii of 50\arcsec\ and 100\arcsec, respectively. 
Ancillary response files (ARFs) and redistribution matrix files (RMFs) were generated accordingly.
The spectra from individual observations were then co-added to produce a combined spectrum for each source by using the SAS task \emph{epicspeccombine}. The combined spectra were adaptively binned to achieve a signal-to-noise ratio (S/N) greater than two per energy bin.

Since our sources were selected from the eRASS1 catalog, we have a prior expectation, supported by Sect.~\ref{sec:hamstar} and Fig.~\ref{fig:xmmgaia}, that the majority ($\gtrsim$70\%) are coronal sources identified by cross-identification in the \textit{HamStar} catalog. These sources are expected to exhibit thermal plasma spectra, while a subset of other sources may instead show non-thermal spectra.
To account for both possibilities, we adopted a phenomenological spectral modeling approach using the Python interface to XSPEC \citep{gordon2021}. 
Each spectrum was independently fit with two alternative models: an absorbed thermal plasma model (\texttt{tbabs}$\times$\texttt{apec}) and an absorbed power-law model (\texttt{tbabs}$\times$\texttt{powerlw}). Both models have a small number of free parameters, allowing us to do a direct comparison of the absolute chi-squared ($\chi^2$) values under two models.
The power-law model was adopted only if it yielded a significantly better fit, defined by a decrease in chi-squared of $\Delta \chi^2 > 9.0 $ compared to the thermal model. This threshold corresponds to approximately 3$\sigma$ of the chi-squared distribution with equal degrees of freedom, indicating strong statistical preference for the power-law model. In all other cases, we adopted the thermal model given its physical relevance for stellar coronal emission and the prior expectation for coronal sources.

For all remaining sources, we used the absorbed thermal plasma model.
The Galactic absorption was modeled using the \texttt{tbabs} component \citep{2000ApJ...542..914W}. The \texttt{apec}\footnote{https://heasarc.gsfc.nasa.gov/xanadu/xspec/manual/node134.html} component represents emission from an optically thin, collisionally ionized plasma and is characterized by three parameters: the plasma temperature ($kT$), the global metal abundance ($Z$), and the emission measure (EM). The emission measure is proportional to the integral of the squared electron density over the emitting volume and is derived from the model normalization, assuming a known distance to the source \citep{Magaudda2022}.

Each spectrum was fit with either a single-temperature model, \texttt{tbabs}$\times$\texttt{apec}, or a two-temperature model, \texttt{tbabs}$\times$(\texttt{apec}+\texttt{apec}), depending on the quality of the fit.
In particular, 
if the single-temperature model yielded a reduced chi-squared $\chi^2_{\rm red} > 1.5$, we adopted a two-temperature model to improve the fit. For these two-component fits, we also derived a mean coronal temperature ($T_{\rm mean}$), calculated as the emission measure–weighted average of the component temperatures:
\begin{equation}
T_{\rm mean} = \frac{\sum (\mathrm{EM}_n \cdot T_n)}{\sum \mathrm{EM}_n},
\end{equation}
where $n = 1, 2$ corresponds to the two components in the best-fitting model.

The best-fitting parameters, including $\Gamma$ or $T_{\rm mean}$ where applicable, are presented in Table~\ref{tab:specsrc}. For sources with distance measurements, we also report the unabsorbed X-ray luminosity in the 0.2--10~keV band based on the best-fit spectral model.

\begin{figure}
  \centering
   \includegraphics[width=\hsize]{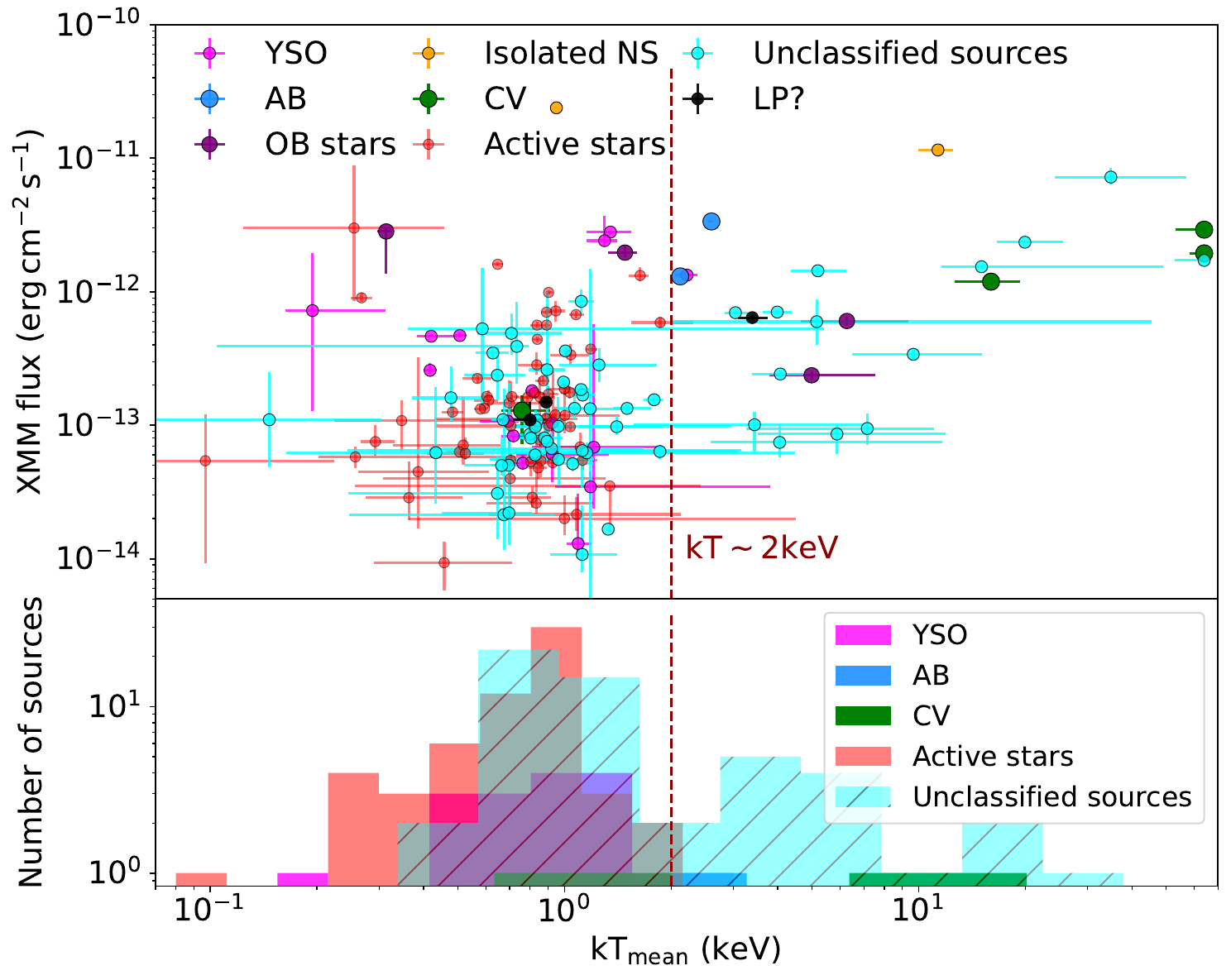}
      \caption{\xmm flux and thermal temperature distributions. 
      Top: 0.2--10~keV unabsorbed X-ray flux versus APEC best-fit temperature ($kT$) for different source classes. 
      Bottom: Histogram of the best-fit mean temperatures for each category.}
       \label{fig:kT_scatter}
\end{figure}

\section{Classification of the eRASS1 sources}
\label{sec:class}
\subsection{Classification methodology}

With reliable \gaia counterparts identified for the majority of sources, we classified all 158 eRASS1 sources with available \textit{XMM-Newton} spectra based on a combination of optical photometry and X-ray spectral properties. 
While 17 sources have \gaia associations considered as ``Low-confidence match'' (i.e., with a false-match probability exceeding 5\%), this does not significantly impact our classification or conclusions. As detailed below, 11 of these sources already have secure identifications in SIMBAD, and the remaining 6 are classified primarily based on their distinct X-ray spectral characteristics. Therefore, the inclusion of these low-confidence matches does not compromise the robustness of our overall classification.

\subsubsection{X-ray spectral properties of SIMBAD-classified sources}
\label{sec:simbadclass}

We first cross-matched the 158 X-ray sources with the SIMBAD database and identified counterparts for 99 of them. These include:

\begin{itemize}
    \item 61 classified as ``Star'', ``Rotating Variable'', ``Eclipsing Binary'' — all interpreted as coronally active stars;
    \item 16 classified as YSOs or YSO candidates, also considered as active stars;
    \item 3 Long Period Variable Candidates (LP?)
    \item 4 cataclysmic variables (CVs);
    \item 2 active binaries (RS CVn systems);
    \item 4 massive OB stars (3 Be stars, 1 Wolf-Rayet star);
    \item 2 isolated neutron stars (including 1 X-ray pulsar);
    \item 4 low-mass X-ray binaries (LMXBs);
    \item 2 high-mass X-ray binaries (HMXBs);
    \item 1 active galactic nucleus (AGN).
\end{itemize}
The remaining 59 sources lack definitive classifications. We consider the 99 SIMBAD-identified sources as a reference sample and plot their mean coronal temperature ($kT_{\rm mean}$) versus X-ray flux in Fig.~\ref{fig:kT_scatter}.

We find that nearly all coronally active stars (and YSOs, hereafter referred to as ``Active stars'' in figures for simplicity) have $kT_{\rm mean} \lesssim 2$~keV, suggesting a practical threshold to distinguish coronal stars from others. 
Based on this, we conservatively assign a temporary classification of active stars to all unclassified sources with $kT_{\rm mean} \lesssim 2$~keV.
However, we note that some massive OB stars in the SIMBAD sample also fall below this temperature threshold, and therefore misclassification is possible (see detailed discussion below in Sect. \ref{sec:cmd}).
Additionally, although both the two confirmed ABs exhibit $kT_{\rm mean} > 2$~keV, we expect some ABs may also lie below 2~keV, and cannot be reliably distinguished from other coronal sources using spectral and photometric information alone.

\subsubsection{Distinguishing compact objects by the ``X-ray Main Sequence'' diagram}
\label{sec:xrayms}

To further distinguish accreting compact objects from coronal sources with hard spectra, i.e., the unclassified sources with $kT_{\rm mean} > 2$~keV in Fig.~\ref{fig:kT_scatter}, we adopt the ``X-ray Main Sequence'' diagram proposed by \citet{Rodriguez2024}. This diagnostic separates populations based on \gaia optical color (BP--RP) and the X-ray-to-optical flux ratio ($F_{\rm X}/F_{\rm opt}$). The empirical boundary is given by:
\begin{equation}
\log_{10} \left( \frac{F_{\rm X}}{F_{\rm opt}} \right) = \mathrm{BP} - \mathrm{RP} - 3.5.
\end{equation}
X-ray fluxes are taken from our best-fit models (Sect.~\ref{sec:spec}) and corrected for absorption using the corresponding best-fit column densities ($N_{\rm H}$). \gaia magnitudes were extinction-corrected using the 3D dust map from \citet{Lallement2019}. When available, distances were adopted from \texttt{distance\_gspphot} in Gaia DR3; otherwise, inverse parallax was used if \texttt{parallax\_over\_error} > 3.5.
The extinction corrections applied were based on \citet{Wang2019}, using the relations:
$A_{\rm V} = 3.16 \times E(B-V)$,
$A_{\rm G} = 0.789 \times A_V$,
$A_{\rm BP} = 1.002 \times A_V$,  
$A_{\rm RP} = 0.589 \times A_V$.

The six known compact objects with \gaia counterparts (4 CVs and 2 XRBs) all lie above the empirical cut, validating its effectiveness. 
Among the unclassified sources, 9 lie above the cut. Of these, 6 with $kT$ > 2 keV are classified as CV candidates due to their elevated $F_{\rm X}/F_{\rm opt}$ ratios and hard X-ray spectra, consistent with the known CVs.

\begin{figure}
  \centering
   \includegraphics[width=\hsize]{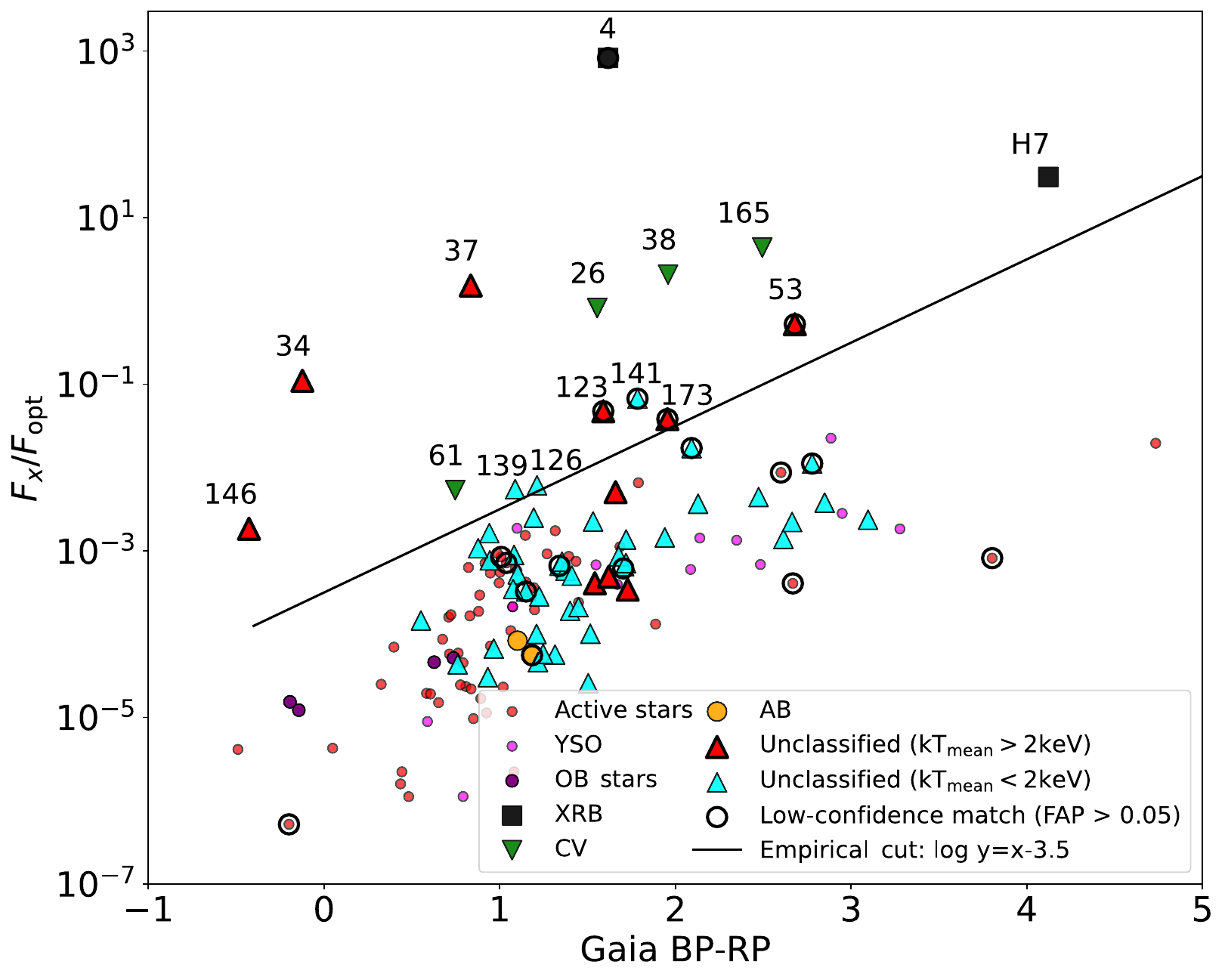}
      \caption[width=\hsize]{“X-ray Main Sequence” diagram (X-ray-to-optical flux ratio ($F_{\rm X}/F_{\rm opt}$) against \gaia BP-RP color). The solid line shows the empirical boundary separating accreting compact objects from coronal sources. Labels indicate source beyond the empirical cut. Low-confidence match are indicated by open black circles.}
       \label{fig:xraymain}
\end{figure}

The remaining four unclassified sources with $kT_{\rm mean} > 2$~keV, represented by red triangles in Fig.~\ref{fig:xraymain} but lying below the empirical flux threshold, are X-ray faint and thus unlikely to be accreting systems. They are most likely ABs or OB stars based on the spectral properties of the confirmed sample in Fig.~\ref{fig:kT_scatter}.
The characterization of them will be discussed further based on their optical colors (see Sect. \ref{sec:cmd} and Sect. \ref{sec:coronal}).

\subsubsection{\gaia color–magnitude diagram}
\label{sec:cmd}
\begin{figure}
  \centering
  \includegraphics[width=1.02\hsize]{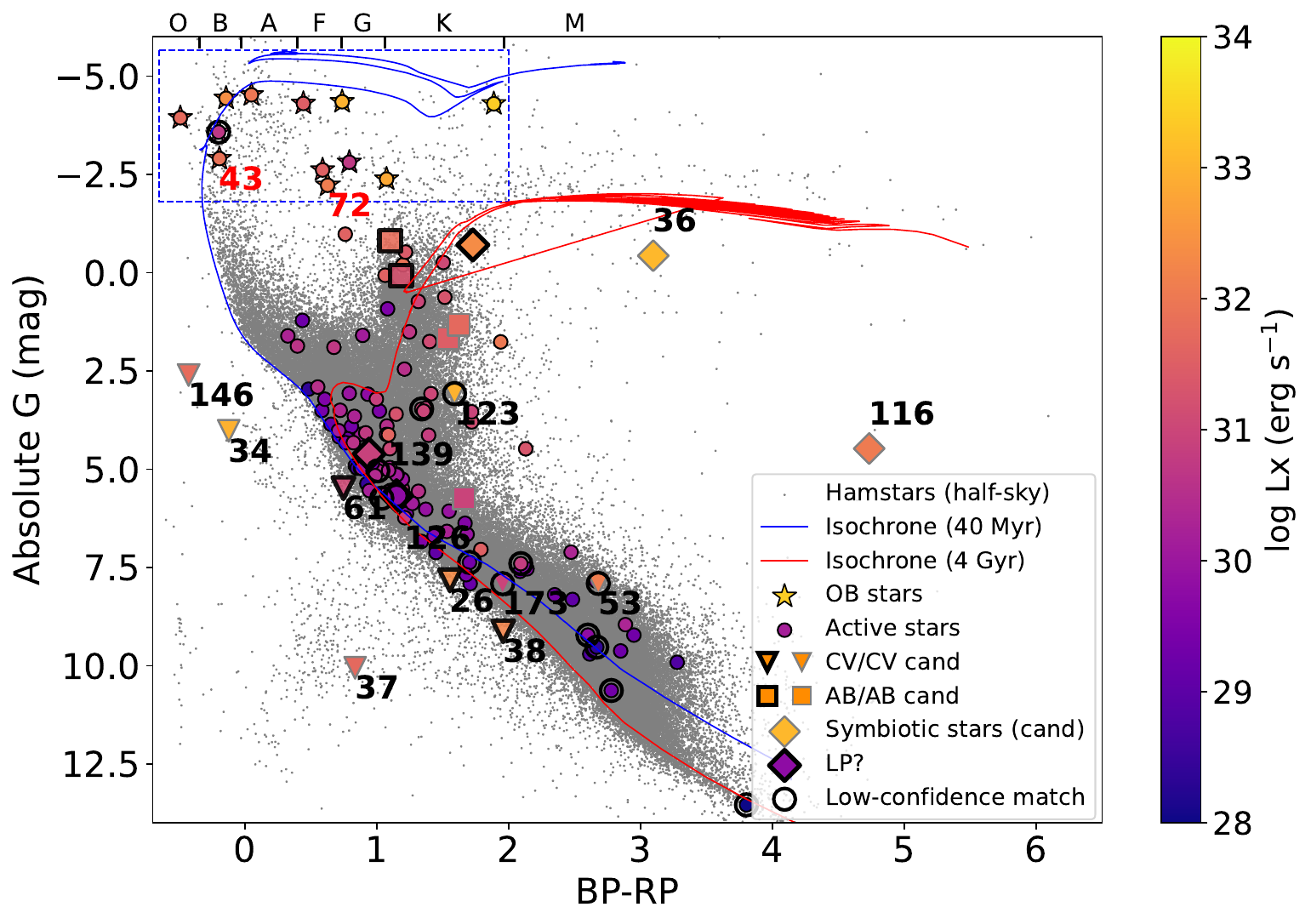}
  \caption{
    \gaia Hertzsprung–Russell diagram for sources with distance estimates. The x-axis shows the \gaia BP-RP color, and the y-axis shows the absolute $G$-band magnitude. The color bar indicates the unabsorbed 0.2--10~keV X-ray luminosity. 
    Overplotted are PARSEC isochrones for stellar populations of 40 Myr (blue) and 4 Gyr (red). Sources above the empirical cut in the ``X-ray Main Sequence'' diagram are labeled with black source IDs, while two $\gamma$~Cas candidates are marked in red. Two extremely ``red'' outliers are marked as SySt candidates. Low-confidence match are shown as open black circles.
  }
  \label{fig:absG_BPRP}
\end{figure}

For sources with distance measurements from \gaia, we construct the Hertzsprung–Russell (HR) diagram using extinction-corrected BP--RP colors and absolute $G$-band magnitudes. Figure~\ref{fig:absG_BPRP} displays the HR diagram, with X-ray luminosity (from unabsorbed 0.2--10~keV spectral fits) encoded in color. For comparison, we overlay PARSEC stellar isochrones corresponding to ages of 40 Myr and 4 Gyr, representing young and old stellar populations, respectively \citep{Bressan2012}.

A prominent group of 12 sources occupies the bright, blue region of the diagram (approximately $M_G = -5$ to $-2$), consistent with massive early-type O and B stars. These sources also exhibit relatively high X-ray luminosities in the range \(L_{\rm X} \sim 10^{31{-}33}~\mathrm{erg~s^{-1}}\). In contrast to late-type stars, whose X-ray emission originates from magnetically heated coronae, OB stars produce X-rays through embedded wind shocks in their radiatively driven stellar outflows.

Despite applying extinction corrections based on 3D dust maps \citep{Lallement2019}, these sources identified as massive OB stars still appear significantly redder than expected in the \gaia color-magnitude diagram.
This discrepancy may be attributed to several factors. First, the extinction corrections themselves are subject to uncertainties, particularly along sightlines with complex or non-uniform dust structures. In regions of high or patchy extinction, the adopted $A_V$ values from 3D maps may underestimate the total line-of-sight reddening, especially if the stars lie behind concentrated dust clouds or at distances where the dust model becomes less constrained.
Additionally, OB stars embedded in or near star-forming regions may be affected by local circumstellar or nebulous material, causing their colors to remain anomalously red even after standard extinction correction procedures.

Two exceptional outliers with extremely red colors and high optical luminosities significantly above the main sequence are identified as candidate symbiotic stars (Src 36 and Src 116), a rare class of interacting binaries where a compact object (typically a white dwarf or neutron star) accretes material from a red giant companion. Source properties of these two are discussed further in Sect. \ref{sec:syst}.
Finally, sources lying above the empirical cut in the ``X-ray Main Sequence'' diagram (Sect.~\ref{sec:xrayms}) are annotated in the figure with black source IDs, these along with candidates for ABs, and those sources without \gaia counterparts are discussed in detail in the following sections.

\begin{figure}
  \centering
   \includegraphics[width=0.95\hsize]{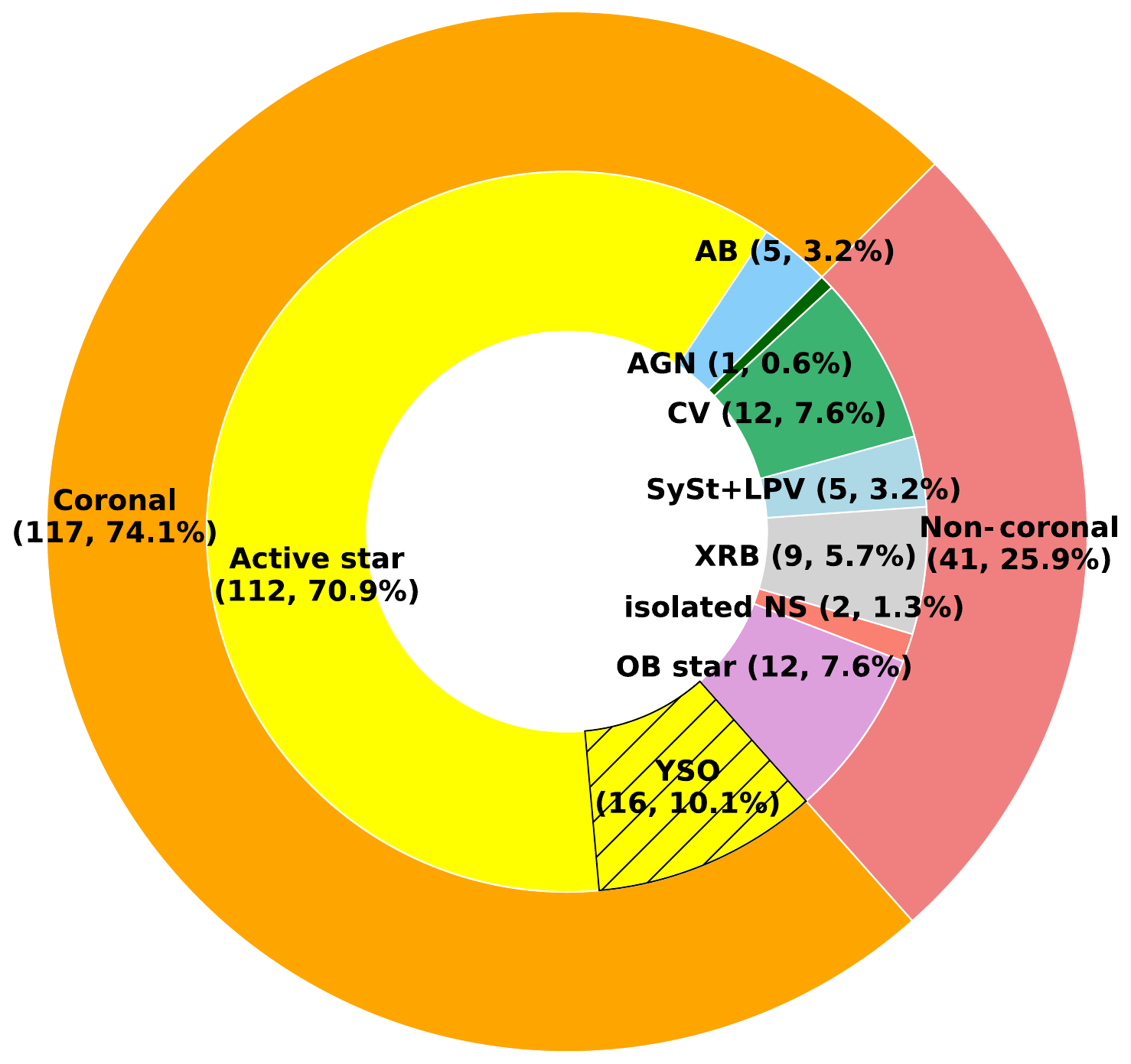}
      \caption[width=\hsize]{Source classification of 158 XMM-Newton-detected eRASS1 sources.} 
       \label{fig:srcclass}
\end{figure}

In summary, we identify 112 coronally active stars (including 16 YSO), 5 ABs, 12 early-type stars (hereafter OB stars), 12 CVs, 2 symbiotic stars and 3 LPVs, 7 LMXBs, 2 HMXBs, 2 isolated neutron stars and 1 AGN, summarized in Fig.~\ref{fig:srcclass}.

\subsection{Coronal sources}
\label{sec:coronal}
Of the 158 X-ray sources in our sample, 120 are classified as coronal emitters based on a combination of reliable \gaia\ photometry (when available and of high confidence) and their soft thermal X-ray spectra. For sources lacking reliable \gaia\ counterparts, the classification relies solely on X-ray spectral properties.

These sources primarily include active single stars (including YSOs) and ABs.
Among the more luminous or harder X-ray sources, ABs are the most likely classification. 
ABs typically consist of a slightly evolved subgiant and a solar- or late-type companion. Their orbital periods range from several hours to tens of days, inducing rapid stellar rotation and enhanced magnetic activity \citep{Walter1981}. This results in strong coronal X-ray emission, typically with X-ray luminosities of $L_{\rm X} \sim 10^{28}$–$10^{32}~\rm erg~s^{-1}$ \citep{Audard2003,Sazonov2006}. 
Due to their abundance and relatively faint X-ray fluxes, ABs are recognized as major contributors to the Galactic Ridge X-ray Emission \citep{Revnivtsev2006}.

However, identifying ABs based solely on spectral hardness and optical photometry remains difficult, especially when distinguishing them from active single stars. 
Nevertheless, according to the Simbad database, two known RS CVn-type systems (Src 93 and Src 153) are identified in our sample. 
These are composed of a subgiant or giant primary with a tidally locked late-type companion, and exhibit elevated X-ray luminosities ($L_{\rm X} \sim 10^{31}$–$10^{32}~\rm erg~s^{-1}$) due to intense magnetic activity. 

In addition to the sources discussed above, three additional objects (Src~68, Src~80, Src~104) exhibit hard X-ray spectra with $kT > 2$~keV, suggesting they are likely ABs. Their elevated plasma temperatures make them unlikely to be active single stars (Fig.~\ref{fig:kT_scatter}). Furthermore, their relatively low X-ray-to-optical flux ratios argue against an interpretation as accreting compact objects.
Their locations on the color–magnitude diagram (CMD; Fig.~\ref{fig:absG_BPRP}) are also inconsistent with the properties expected for massive OB stars. Based on this multi-wavelength evidence, we classify these three sources as candidate ABs, marked with gray-edged boxes in Fig.~\ref{fig:absG_BPRP}.

\subsection{Non-coronal sources}

\subsubsection{Early-type stars (binaries)}
Early-type O and B stars are prominent X-ray emitters in young stellar populations. Unlike late-type stars, whose X-ray emission is powered by magnetically heated coronae, OB stars generate X-rays through embedded shocks within their radiatively driven stellar winds. These processes typically produce soft thermal spectra with plasma temperatures of $kT \sim 0.5{-}1~\mathrm{keV} $, and X-ray luminosities reaching $L_{\rm X}\sim 10^{31-33}~\mathrm{erg~s^{-1}}$ \citep{2009A&A...506.1055N}, especially in binary systems where wind–wind collisions can significantly enhance the emission \citep{2021MNRAS.503..715C,2022hxga.book..108R}.

We identify a group of 12 sources in the CMD (Fig.~\ref{fig:absG_BPRP}) consistent with massive OB stars, roughly spanning absolute $G$-band magnitudes of \( M_G = -5 \) to \( -2 \), and exhibit X-ray luminosities in the range \( L_{\rm X} \sim 10^{31{-}33}~\mathrm{erg~s^{-1}} \). 
Their positions in the CMD and X-ray properties are consistent with wind-driven emission from young, massive early-type stars.

Among this group, two sources (Src 43 and Src 72; highlighted with red labels in Fig.~\ref{fig:absG_BPRP}) are noteworthy for their unusually hard X-ray spectra, as presented in Fig.~\ref{fig:yCASspec}. 
Spectroscopic observations confirm that both have classical Be star optical counterparts—early B-type stars that exhibit Balmer-line emission from circumstellar decretion disks. 
Their X-ray spectra require hot thermal components with plasma temperatures of \( kT \sim 5{-}10~\mathrm{keV} \), far exceeding those expected from typical OB stars \citep{2009A&A...506.1055N}.

With X-ray luminosities around \( L_{\rm X}\sim 10^{32}~\mathrm{erg~s^{-1}} \), these sources exhibit a combination of characteristics consistent with the rare class of $\gamma$~Cassiopeiae (``$\gamma$~Cas'') analogs \citep{2018A&A...619A.148N}. 
However, due to the limited quality of the available X-ray spectra,  we cannot conclusively detect key spectral diagnostics, such as an Fe~\textsc{xxv} line, nor can we robustly distinguish between a hot thermal plasma and a power-law continuum.
Therefore, we cannot rule out other plausible interpretations. For instance, quiescent Be/neutron star binaries can also display comparable hard spectra and X-ray luminosities \citep{Elshamouty2016}. Distinguishing between a $\gamma$~Cas analog and a quiescent Be/neutron star binary in these sources would require higher-quality X-ray spectra and long-term X-ray monitoring to search for transient behavior.

In summary, we identify 12 sources in our sample whose X-ray emission is consistent with wind-powered processes in massive OB stars. We classify these objects as non-coronal sources, emphasizing that their X-ray emission arises through fundamentally different mechanisms compared to the magnetically heated coronae of late-type stars and is also distinct from accretion-powered sources such as CVs and LMXBs.

\subsubsection{Cataclysmic variables}

Our sample includes four known CVs listed in the SIMBAD database: AX J1740.3-2904, AX J1740.1-2847, 4XMM J173058.9-350812, and V478 Sco. The first three were also identified with periodicities in our earlier work using the XMM Heritage survey \citep{Mondal24}.

In addition, we identify eight new CV candidates based on their elevated X-ray luminosities relative to the ``X-ray main sequence'' and/or their hard X-ray spectra, both of which are indicative of accreting compact binaries rather than coronal sources. 
Their spectra plots with best-fit models are presented in Fig.~\ref{fig:cvspec}. 
It is worth noting that Src 34 yields a potential super-Solar abundance at $1.7^{+1.0}_{-0.7} ~Z_{\odot}$. Such high abundances are unusual for CVs, where plasma metallicities are typically sub-Solar to Solar \citep{Yuasa2010}. Similar cases have occasionally been reported in a few systems \citep{Harrison2016}, possibly reflecting peculiar donor compositions or accretion histories. Given the modest S/N of our data and potential model degeneracies, this result should be interpreted with caution and requires confirmation with deeper observations.

Two of these candidates (Src~138 and Src~144) lack \gaia counterparts, suggesting either large distances or optically faint donor stars. 
Both exhibit high plasma temperatures ($kT > 5$~keV), consistent with emission from the accretion column.
Src~138 also shows prominent Fe~\textsc{xxv} and Fe~\textsc{xxvi} emission lines at 6.7 and 7.0 keV, respectively, strongly supporting its classification as a CV. 
Although Src~144 has fewer counts and lacks identifiable Fe line features, we tentatively classify Src~144 as a CV candidate, while a quiescent LMXB cannot be ruled out.

Among the remaining six candidates, four sources (Src~34, Src~53, Src~123, and Src~173) have \gaia counterparts with false-match probabilities exceeding 5\%, indicating that their associations are not secure. 
Nevertheless, all four exhibit hard X-ray spectra with $kT > 4$~keV, consistent with the spectral properties of confirmed CVs shown in Fig.~\ref{fig:kT_scatter}.
In addition, Src~34, Src~53, and Src~123 display prominent Fe~\textsc{xxv} emission lines, a characteristic feature of CVs, further supporting this classification. Src~173, on the other hand, lacks clear a Fe line detection due to limited spectral quality.

We also examine the Gaia color–magnitude diagram (Fig.~\ref{fig:absG_BPRP}).
Three candidates (Src~34, Src~37, and Src~146) lie along the white dwarf sequence, lending further support to their identification as CVs.
Interestingly, most known CVs and our additional candidates do not appear on the white dwarf branch in the Gaia CMD. 
This is not unexpected as CVs are interacting binaries where the optical emission might be dominated by the donor star or the accretion disk, rather than the white dwarf itself. Particularly in systems with luminous secondaries or high accretion rates, the observed colors and magnitudes deviate significantly from those of isolated white dwarfs.

In summary, we identify eight new candidate CVs, in addition to the four previously known systems. These candidates are primarily selected based on their locations in the "X-ray Main Sequence" diagram (Fig.~\ref{fig:xraymain}), which serves as an effective diagnostic tool for distinguishing accreting systems from coronal sources.
Further support for their classification comes from their positions in the CMD (Fig.~\ref{fig:absG_BPRP}), where three of the new candidates fall along the white dwarf sequence, consistent with expectations for CVs.
Despite uncertainties or potential mismatches in their \gaia associations, the X-ray spectral properties of these sources, characterized by high plasma temperatures and prominent Fe~\textsc{xxv}/Fe~\textsc{xxvi} line emission, provide additional evidence for their identification as accreting white dwarf systems.

\subsubsection{Symbiotic stars and long period variables}
\label{sec:syst}
Src~36 and Src~116 stand out as clear outliers in the \gaia color-magnitude diagram (Fig.~\ref{fig:absG_BPRP}), exhibiting extremely red colors ($\mathrm{BP{-}RP} > 3$) and luminosities significantly above the main-sequence, consistent with late-type giants \citep{Merc2020,2025arXiv250620505B}.  Their X-ray spectral plots are presented in Fig.~\ref{fig:systspec}. Both sources have  $L_{\rm X} \gtrsim 10^{32}~\mathrm{erg~s^{-1}}$ and display relatively soft X-ray spectra with plasma temperatures of $kT \sim 1$–2~keV, which are notably softer than typical CVs. 
Notably, Src~36 was previously identified as an AGB star based on infrared photometry \citep{AGBerass1}, further supporting it could be a symbiotic star in which a white dwarf or neutron star accretes material from an AGB donor.

We classify these two objects as strong candidates for symbiotic stars (SySt). These systems are characterized by very long orbital periods ($P_{\rm orb} \gtrsim 200$ days), strong optical and infrared emission from the giant star, and diverse X-ray properties that range from soft, thermal spectra to harder, more complex emission depending on the accretion regime and the presence of winds or shocks \citep{Symbiotic}. 

In addition to these two symbiotic candidates, our sample contains three sources identified as Long Period Variables (LPVs) by \citet{Lebzelter2023}. The presence of X-ray emission in these LPVs, a phenomenon not inherent to these single stars, is probably due to wind accretion onto an unseen white dwarf companion, as also suggested by \citet{Schmitt2024}. Given this shared mechanism, we group the LPV candidates with the SySts. However, it is important to note that without UV data to confirm the presence of a hot companion, the exact origin of their X-ray emission remains uncertain.

\subsection{Isolated neutron stars}

Our sample includes two sources classified as isolated neutron stars (NS). And their X-ray spectra are shown in Fig.~\ref{fig:nsspec}.
Src~7 is a central compact object (CCO) located within the supernova remnant HESS~J1731$-$347. This source has been proposed as the lightest neutron star ever observed, with estimated $M = 0.77^{+0.20}
_{-0.17}~\rm M_\odot$ \citep{CCO22}. It exhibits a soft, thermal X-ray spectrum well-described by a blackbody model with a temperature of $kT \sim 0.5$~keV, and an unabsorbed luminosity of $L_{\rm X} \sim 2 \times 10^{34}$~erg~s$^{-1}$.
Another isolated NS in our sample is Src~H2, corresponding to PSR~J1747$-$2958—an X-ray pulsar discovered within the ``Mouse'' radio nebula (G359.23$-$0.82), with a spin period of $P = 98$~ms \citep{PSRJ1747}. It is identified as a ``Hard-only'' source, exhibiting a hard X-ray spectrum ($\Gamma \sim 1.77$) and a luminosity of $L_{\rm X} \sim 9 \times 10^{33}$~erg~s$^{-1}$.

\subsection{LMXBs and HMXBs}

Four LMXBs (Src~1, Src~2, Src~4, and Src~5) and two HMXBs (H1, H7) are classified based on entries in the \texttt{SIMBAD} database, with additional support from their X-ray spectral properties, despite the lack of confirmed optical counterparts.

In addition, we identify three new LMXB candidates and present their spectral plots in Fig.~\ref{fig:lmxbspec}.
Src~6, situated in the globular cluster Liller~1, shows a hard X-ray spectrum ($\Gamma \sim 1.34$) and an X-ray luminosity of $L_{\rm X} \sim 5 \times 10^{33}$~erg~s$^{-1}$. This source was also reported as a new X-ray transient by \textit{Swift} \citep{ATel2018}. Assuming a cluster distance of 8.2~kpc \citep{Harris1996}, the peak 0.5--10~keV luminosity reaches $\sim 5.5 \times 10^{35}$~erg~s$^{-1}$ \citep{ATel2018}.

Src~32 lacks a \gaia counterpart and therefore cannot be reliably placed on the ``X-ray Main Sequence'' diagram for classification. However, its X-ray spectrum is exceptionally hard, characterized by a power-law photon index of $\Gamma \sim 1.5$, and shows no detectable Fe~\textsc{xxv} or Fe~\textsc{xxvi} emission lines. 
Spectral modeling reveals substantial absorption, with a column density of $N_{\rm H} \sim 2.5 \times 10^{22}~\mathrm{cm}^{-2}$. According to the 3D dust extinction map from \citet{Lallement2019}, we estimate a minimum distance of $\sim$3~kpc for this source. At this distance, the implied X-ray luminosity exceeds $L_{\rm X} \approx 3 \times 10^{33}\,(d/3~\mathrm{kpc})^2~\mathrm{erg\,s}^{-1}$. Taken together, the high intrinsic absorption, hard spectrum with no Fe line features, and high luminosity strongly suggest that Src~32 is a LMXB candidate.

Src~H5 also lacks a \gaia counterpart and exhibits a hard X-ray spectrum ($\rm \Gamma\sim 1.43$) with no discernible Fe~\textsc{xxv} or Fe~\textsc{xxvi} emission lines. This source was initially detected as a transient by {\it Swift}-XRT in May 2012 \citep{2012ATel.4111....1B}, reaching a luminosity of $\sim 3.8 \times 10^{36}~\text{erg s}^{-1}$ (assuming 8~kpc). In our observations, we found the source to be approximately 20 times fainter, with a luminosity of $\sim 1.9 \times 10^{35}~\text{erg s}^{-1}$ (also for 8~kpc). \citet{2012ATel.4111....1B} ruled out the possibility of it being a Supergiant Fast X-ray Transient (SFXT) due to the absence of infrared stellar sources within its error circle. Based on these findings, we tentatively classify Src~H5 as an LMXB.

\subsection{Background AGN}

Src~3 is the only AGN identified in our sample. It has been classified as a Seyfert 2 galaxy with a redshift-estimated distance of approximately 93~Mpc \citep{2022ApJS..261....2K}, corresponding to an X-ray luminosity of $L_{\rm X} \sim 6.2 \times 10^{43}~\mathrm{erg~s^{-1}}$. 
According to \citet{erass1AGN}, a total of 2547 eRASS1 sources are classified as reliable AGN candidates across the western Galactic hemisphere. Given our survey's sky coverage of roughly $20~\mathrm{deg}^2$, we expect $\sim$1.23 AGN to fall within this region, which is consistent with the single AGN identified in our sample.

\section{Discussion}
\label{sec:discuss}

\subsection{An empirical hardness ratio cut to identify accreting compact objects in eRASS1}

Due to limited photon statistics, the majority of eRASS1 sources lacks high-quality X-ray spectra. Consequently, the hardness ratio (HR) serves as a practical proxy for distinguishing different classes of X-ray emitters.
Our sample, which includes tentative classifications based on \xmm spectra, provides a representative subset of the X-ray source population along the Galactic plane.

\begin{figure}
  \centering
   \includegraphics[width=\hsize]{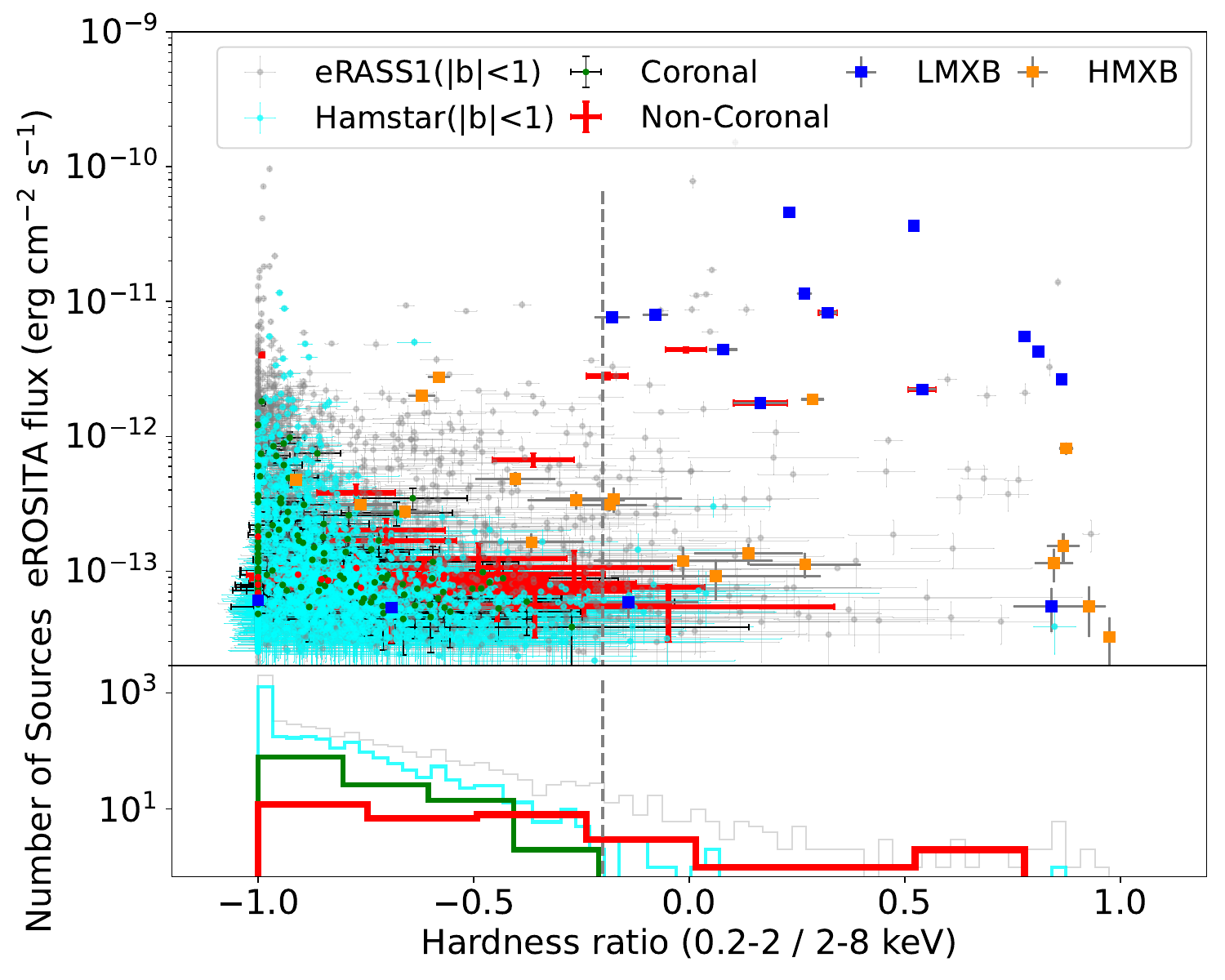}
      \caption[width=\hsize]{
     Hardness ratio versus flux for eRASS1 Galactic plane sources ($|b| < 1\degr$). Upper panel: The hardness ratio against the 0.2--8 keV flux for eRASS1 sources within $|b|< 1\degr$. The vertical dashed line represents an empirical cut at HR=-0.2. Known Galactic LMXBs and HMXBs from \citet{LMXBcats,HMXBcats} are indicated by blue and orange squares, respectively. Lower panel: histograms of HR for different source categories. 
     }
       \label{fig:HRerass1}
\end{figure}

We classify the sources into two broad categories: coronal sources, including coronally active stars and ABs; and non-coronal sources, comprising wind-powered massive OB stars, isolated NSs and accreting compact objects including CVs, SySts, LMXBs, and AGNs. 
Using hardness ratios measured by eROSITA, defined here as $\mathrm{HR} = (\rm H - S)/(H + S)$, where S and H represent count rates in the 0.2--2~keV and 2--8~keV bands, respectively, we find that a threshold at $\mathrm{HR} > -0.2$ effectively separates non-coronal (red cross) from coronal sources (black cross with green dots) as shown in Fig.~\ref{fig:HRerass1}.

To generalize this empirical cut, we examine all eRASS1 sources located along the Galactic plane (|$b$|~$< 1\degr$). As shown in Fig.~\ref{fig:HRerass1}, out of 4774 such sources, 2757 are identified as coronal using the \textit{HamStar} classifier (excluding those flagged with \texttt{OB\_STAR == ``TRUE''}). Only 13 of these coronal sources exhibit $\mathrm{HR} > -0.2$, corresponding to an efficiency of $\sim$99.6\% in rejecting coronal contaminants. 
Among all 4774 sources, 149 satisfy the $\mathrm{HR} > -0.2$ criterion. After accounting for the 13 coronal contaminants, this yields a selection purity of approximately 91\%, strongly suggesting that the majority of HR-selected sources are non-coronal sources. This minor contamination can be attributed to rare cases where very hot and heavily absorbed coronal sources exhibit hardness ratios above our -0.2 threshold. For example, in an extreme case of a coronal source with $\rm kT=2.0$ keV and $L_{\rm X} \sim 10^{32}$ erg s$^{-1}$, an absorption of $N_{\rm H} = 3.5 \times 10^{21}$ cm$^{-2}$ would shift its HR upward. Such a source would remain detectable within $\lesssim 1.5$ kpc given the eRASS1 sensitivity limit.

\begin{figure}[htbp]
  \centering
   \includegraphics[width=\hsize]{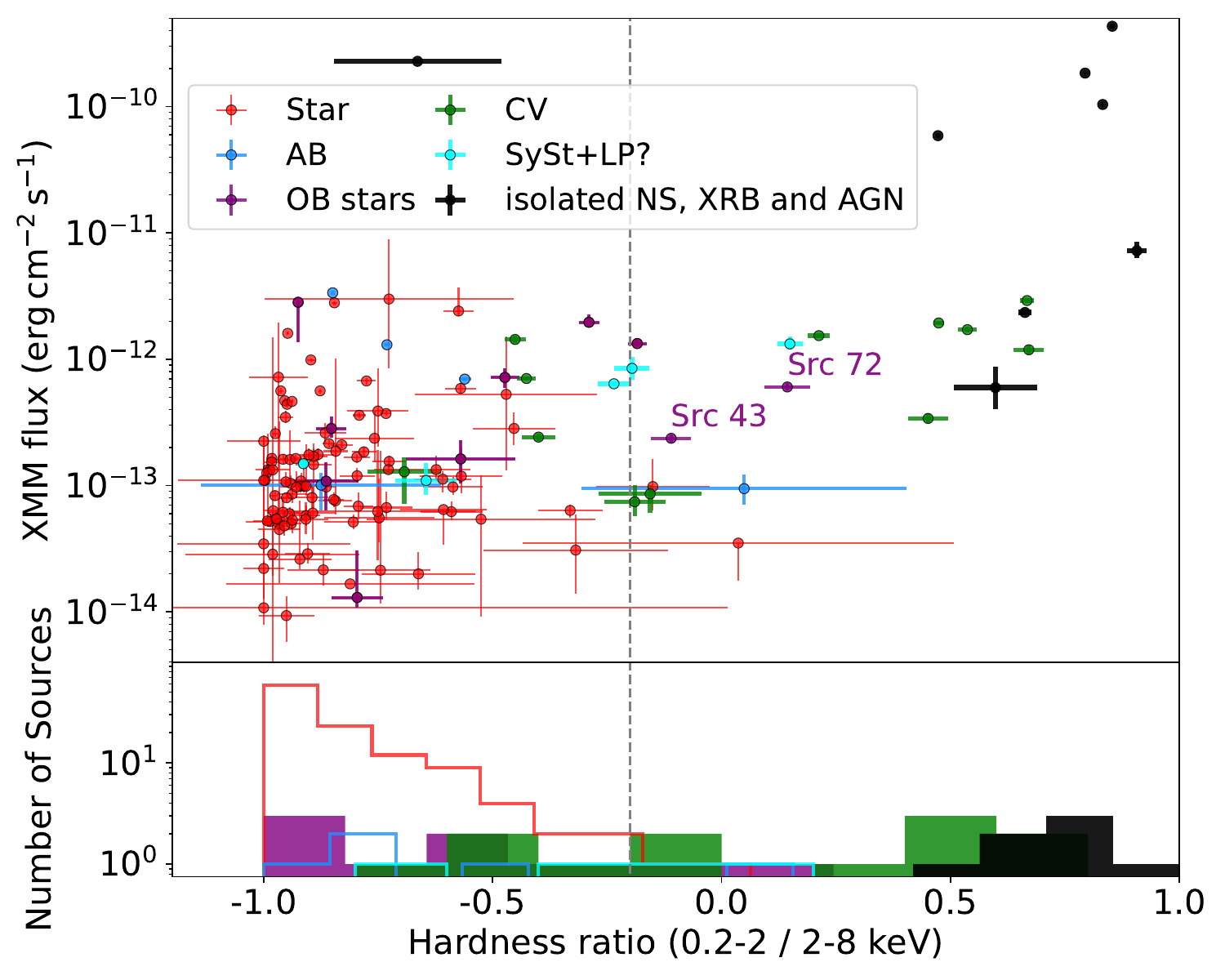}
      \caption[width=\hsize]{Hardness ratio versus 0.2--10 keV flux from XMM spectral fitting. Upper panel: the hardness ratio against 0.2--8 keV flux. The vertical dashed line represents the empirical cut at HR=– 0.2.
      Lower panel: Hardness ratio distributions for different source classes.}
       \label{fig:HRxmm}
\end{figure}

Further support for this HR cut comes from \xmm data, where better-constrained hardness ratios are available (Fig.~\ref{fig:HRxmm}). 
Among 117 sources classified as coronal via their XMM spectra+\gaia cross-match, only 3 exceed $\mathrm{HR}>-0.2$. In contrast, 20 of 41 non-coronal sources fall above this threshold, resulting in a selection purity of $\sim$87\%. This is consistent with the result derived from the eRASS1 sample and further reinforces the reliability of the HR cut. 
The minor discrepancy can be attributed to the $\sim$91.5\% completeness of the \textit{HamStar} classification scheme \citep{hamstar}. 
Notably, all but one of the LMXBs in the labeled sample lie above the threshold, highlighting the effectiveness of the HR cut in identifying this population. The selection also successfully recovers over half of the CVs and a subset of wind-powered OB stars. Residual contamination primarily stems from a few active stars and ABs with intrinsically hard X-ray spectra and low photon statistics.

According to \citet{erass1AGN}, 2547 eRASS1 sources are reliable AGN candidates. 
Assuming a uniform sky distribution and accounting for the effects of Galactic extinction, no more than $\sim$44 AGNs are expected within |$b$|~$<1\degr$. Even under the conservative assumption that all of these AGNs satisfy $\mathrm{HR} > -0.2$, this still leaves $\sim$90 sources as OB stars and Galactic accreting systems, such as XRBs and CVs.
Referring to the cataloged Galactic XRBs \citep{LMXBcats, HMXBcats}, we identify 17 known LMXBs and 20 known HMXBs within |$b$|~$< 1\degr$. Of these, 14 LMXBs and 12 HMXBs satisfy the HR cut, supporting the use of $\mathrm{HR} > -0.2$ as a simple yet efficient empirical criterion for isolating accreting compact objects, particularly LMXBs, within the eRASS1 Galactic plane sample.

\subsection{Resolved point source contribution to the GRXE}
\label{sec:grxe_contribution}

\begin{figure*}
  \centering
  \includegraphics[width=\hsize]{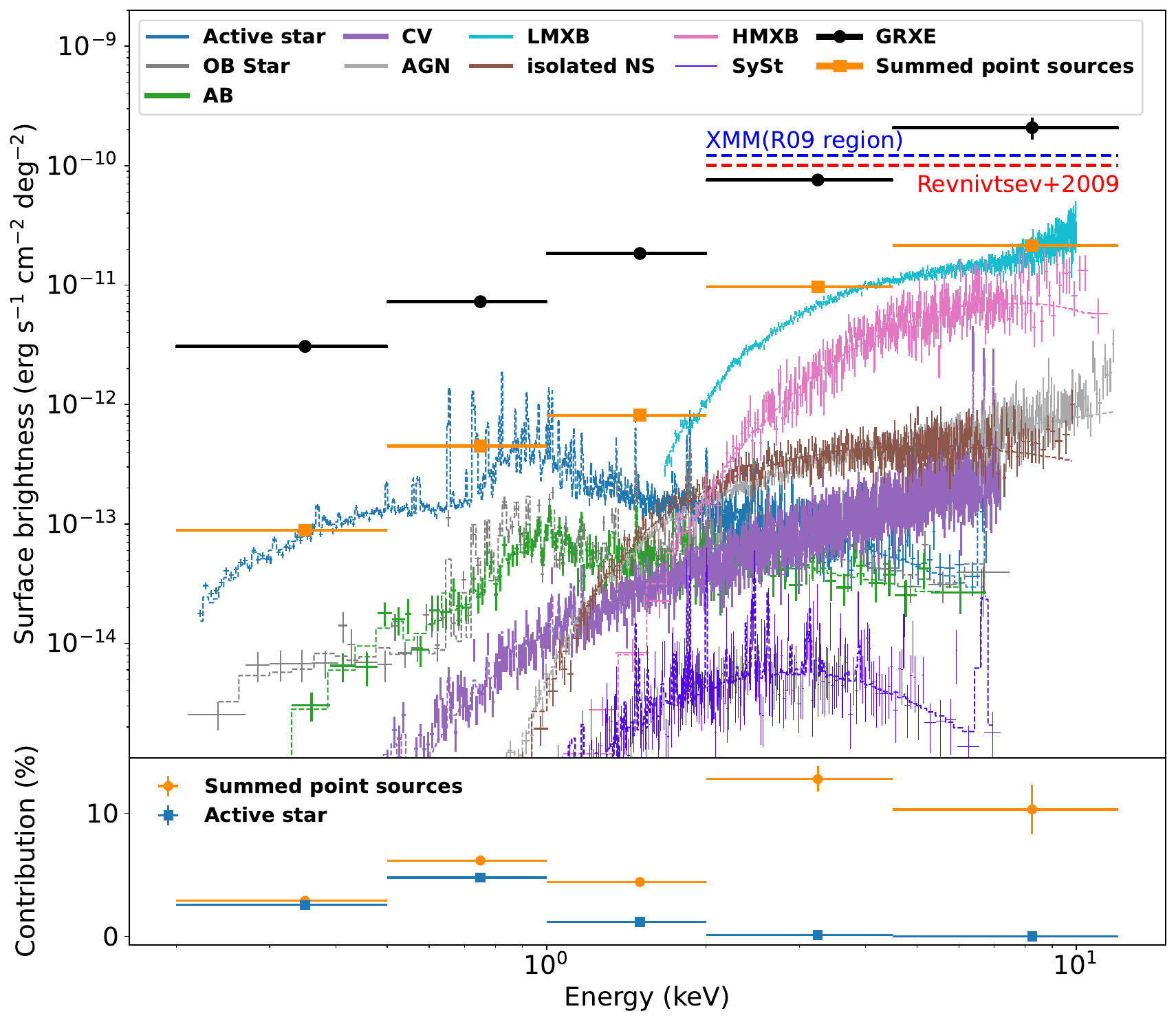}
  \caption{
  Stacked X-ray spectra of classified source populations from the eRASS1 sample (colored lines), compared to the GRXE surface brightness (black points) in five broad energy bands: 0.2--0.5, 0.5--1.0, 1.0--2.0, 2.0--4.5, and 4.5--12.0~keV. The summed spectrum of all classified point sources is shown as orange squares.}
  \label{fig:GRXE_spec}
\end{figure*}

Using data from our \xmm heritage survey, we measure the X-ray surface brightness towards the inner Galactic disk. The total X-ray photon flux is converted into energy flux assuming a thermal plasma with temperature $kT\sim$ 10 keV, representative of the GRXE spectrum observed by {\it Chandra} \citep{Revnivtsev2009}. Surface brightness values are computed in five standard energy bands: 0.2–0.5, 0.5–1.0, 1.0–2.0, 2.0–4.5, and 4.5–12.0 keV, across the western portion of our sky coverage, excluding the central 15\arcmin\ around the Galactic Center.

The total surface brightness of GRXE has been measured previously by other missions.
In the ``Limiting Window'' region ($l^{\rm II} = 0.113\degr$, $b^{\rm II} = -1.424\degr$), \citet{Revnivtsev2009} reported a total X-ray surface brightness of $\rm I_{2-10~keV} = 8.6 \times 10^{-11}~erg~s^{-1}~cm^{-2}~deg^{-2}$ based on a 1~Ms \textit{Chandra} observation. Adopting the same thermal model of $kT\sim$ 10 keV, this corresponds to $\rm I_{2-12~keV}=10^{-10}~erg~s^{-1}~cm^{-2}~deg^{-2}$.
For comparison, our measurements using \textit{XMM-Newton} data yield a slightly higher, yet broadly consistent surface brightness of $\rm I_{2-12~keV} = 1.2 \times 10^{-10}~erg~s^{-1}~cm^{-2}~deg^{-2}$ in the same region.
These values are indicated as red and blue dashed lines in Fig.~\ref{fig:GRXE_spec} for direct comparison.

In addition, we stack the X-ray spectra of each major source class identified in our sample and rebin their photon flux densities into five standard energy bands: 0.2--0.5, 0.5--1.0, 1.0--2.0, 2.0--4.5, and 4.5--12.0~keV (Fig.~\ref{fig:GRXE_spec}).
The eRASS1-selected sample, with a typical sensitivity limit of $\sim 5\times 10^{-14}$~erg~cm$^{-2}$~s$^{-1}$ in the 0.5--2.0 keV band, provides a flux-limited view of the soft Galactic X-ray source population.  At this detection threshold, we find that resolved point sources account for approximately 3.0\%, 6.2\%, 4.5\%, 12.8\% and 10.3\% of the GRXE surface brightness in the  0.2--0.5~keV, 0.5--1.0~keV, 1.0--2.0~keV, 2.0--4.5~keV, 4.5--12.0~keV bands, respectively.
In the soft X-ray regime (below 1 keV), the resolved emission is overwhelmingly dominated by coronally active stars, which account for more than 80\% of the detected point sources. The majority of these sources are foreground objects located within 1 kpc, a finding that supports the long-standing hypothesis that late-type stars are the primary point-source class contributing to the soft component of the GRXE \citep{Revnivtsev2006,Revnivtsev2007}. The resolved composition from this representative region therefore provides a crucial benchmark for the source populations expected from eRASS observing results of the broader Galactic plane.

It worth noting that the observed resolved fraction is subject to significant selection effects. Due to the high level of interstellar absorption towards the inner Galactic disk, the vast majority of our resolved X-ray sources are foreground stars. Thus, the resolved fraction of soft X-ray sources presented here is a lower limit to the intrinsic contribution of stars to the GRXE.
Despite the improved angular resolution and sensitivity of \textit{XMM-Newton}, more than 90\% of the GRXE remains unresolved across all energy bands, implying the presence of a substantial population of faint X-ray sources lying below the eRASS1 flux limit. Besides, previous deep \textit{Chandra} observations \citep{Revnivtsev2009,Hong2012} have shown that over 80\% of the hard (6--7~keV) GRXE can be resolved into point sources, primarily CVs and ABs. However, the limited effective area of eROSITA in the hard X-ray band reduces its sensitivity to such faint, hard-spectrum sources.

Our findings are consistent with previous deep-field surveys, which emphasize that resolving the bulk of the GRXE requires significantly deeper exposures ($\lesssim$$10^{-15}$~erg~cm$^{-2}$~s$^{-1}$) and higher spatial resolution. Nonetheless, our work represents a significant step forward: for the first time, the soft X-ray spectral component of the GRXE has been decomposed into contributions from distinct source classes, enabled by combining X-ray spectroscopy with \gaia-based optical classifications.
Future missions such as {\it AXIS}, with its planned Galactic Survey program, will enable even deeper and more complete resolution of the GRXE.

\section{Summary and conclusions}

\begin{enumerate}
    \item We present a comprehensive classification of 158 X-ray sources towards the inner Galactic disk ($350\degr < l < 360\degr$, $|b| < 1\degr$), by combining eRASS1 detections with deeper \textit{XMM-Newton} observations and \gaia astrometry. Improved X-ray positions from \textit{XMM-Newton} allow us to exclude $\sim3\%$ of false matches in the initial \textit{HamStar} catalog. Additionally, 2\% of the matched \textit{HamStar} sources are classified as non-coronal, consistent with their estimated reliability of 91.5\%.
    
    \item Our classification reveals that approximately 74\% of eRASS1 sources towards the inner disk are coronal sources (active stars and binaries), with X-ray luminosities spanning from $10^{28}$ to $10^{31}~\mathrm{erg~s^{-1}}$. This includes a subset of ABs with $L_{\rm X}$ reaching up to $\sim 10^{32}~\mathrm{erg~s^{-1}}$. The remaining $\sim$26\% of the sources consist of massive OB stars, CVs with $L_{\rm X} \sim 10^{30}$--$10^{33}~\mathrm{erg~s^{-1}}$, isolated NSs, and more luminous sources such as XRBs and background AGNs.

    \item We demonstrate that a simple hardness ratio cut (HR $> -0.2$) efficiently isolates non-coronal sources in the eRASS1 catalog, with a selection purity of $\sim$91\%. This empirical criterion performs well for identifying non-coronal sources in the Galactic plane and is broadly applicable for selecting foreground sources. Applying this cut, we estimate that over 90 eRASS1 sources in the Galactic plane ($|b|<1\degr$), after accounting for AGN contamination, are likely non-coronal.

    \item By stacking the classified eRASS1 sources and comparing their combined X-ray spectra with GRXE measurements, we estimate that $\sim$6\% of the GRXE can be resolved into point sources above the eRASS1 flux limit of $\sim 5 \times 10^{-14}$~erg~cm$^{-2}$~s$^{-1}$ in the 0.5--2.0 keV band, most of which are foreground coronal sources. While the bulk of the soft GRXE remains unresolved, our analysis enables a physical decomposition of its resolved fraction, showing that coronally active stars are the dominant contributors to the soft-band emission.

\end{enumerate}

\section*{Data availability}
  Tables \ref{tab:specsrc} will be available in electronic form at the CDS via anonymous ftp to cdsarc.u-strasbg.fr (130.79.128.5) or via http://cdsweb.u-strasbg.fr/cgi-bin/qcat?J/A+A/.

\begin{acknowledgements}
T.B. and G.P acknowledges financial support from Bando per il Finanziamento della Ricerca Fondamentale 2022 dell’Istituto Nazionale di Astrofisica (INAF): GO Large program and from the Framework per l’Attrazione e il Rafforzamento delle Eccellenze (FARE) per la ricerca in Italia (R20L5S39T9)
G.P. also acknowledges financial support from the European Research Council (ERC) under the European Union's Horizon 2020 research and innovation program HotMilk (grant agreement No. 865637).
SM and MRM acknowledge support from NASA ADAP grants 80NSSC24K0666 and 80NSSC24K0639, respectively.

This work has made use of data from the European Space Agency (ESA) mission
{\it Gaia} (\url{https://www.cosmos.esa.int/gaia}), processed by the {\it Gaia}
Data Processing and Analysis Consortium (DPAC,
\url{https://www.cosmos.esa.int/web/gaia/dpac/consortium}). Funding for the DPAC
has been provided by national institutions, in particular the institutions
participating in the {\it Gaia} Multilateral Agreement.
      
\end{acknowledgements}

\bibliographystyle{aa}
\bibliography{refs.bib}

\begin{appendix} 
\onecolumn
\section{Source catalog with X-ray spectral properties and classifications}
\setlength{\tabcolsep}{2pt}
\renewcommand{\arraystretch}{1.5}
\begin{longtable}{cccccccccccc}
\toprule
ID  & XMM\_RA & XMM\_DEC & SEP\_EX & SEP\_XGAIA & $\rm P_{false}$ & D (kpc) & TYPE &SIMBAD & $\rm$ $\Gamma$ / $\rm kT_{\rm mean}$ & $\chi_{\rm red}^2$ & log10 ($L_{\rm X}$) \\
\midrule
\endfirsthead
\toprule
ID & XMM\_RA & XMM\_DEC & SEP\_EX  & SEP\_XGAIA & $\rm P_{false}$ & D (kpc) & TYPE &SIMBAD & kT & $\chi_{\rm red}^2$ & log10 ($L_{\rm X}$) \\
\midrule
\endhead
\bottomrule
\endfoot
1$^*$ & 264.97436 & -28.49609 & 1.93 & -- & / & $^\dag4.06$ & LMXB & LMXB & $^\ddag 1.32^{+0.01}_{-0.01}$ & 1.52 & $35.65^{+0.00}_{-0.00}$  \\
2$^*$ & 266.52156 & -29.51496 & 0.88 & -- & / & $^\dag8.10$ & LMXB & LMXB & $^\ddag 1.73^{+0.04}_{-0.04}$  & 1.46 & $36.53^{+0.00}_{-0.00}$  \\
3$^*$ & 264.36851 & -29.13426 & 1.69 & 1.45 & 0.10 & $^\dag93190$ & AGN & Seyfert 2 Galaxy & $^\ddag 1.38^{+0.03}_{-0.03}$ & 1.01 & $43.79^{+0.00}_{-0.00}$  \\
4$^*$ & 266.8586 & -30.04508 & 1.45 & 2.61 & 0.09 & -- & LMXB & LMXB & $^\ddag 1.51^{+0.02}_{-0.02}$ & 1.29 & $--^{+--}_{---}$  \\
5$^*$ & 266.85777 & -30.00057 & 3.06 & -- & / & -- & LMXB & LMXB & $^\ddag 1.53^{+0.02}_{-0.02}$ & 1.44 & $--^{+--}_{---}$  \\
6$^*$ & 263.35047 & -33.38964 & 6.33 & 1.46 & 0.07 & $^\dag8.20$ & LMXB? & XrayS & $^\ddag 1.34^{+0.35}_{-0.35}$ & 0.80 & $33.68^{+0.21}_{-0.14}$  \\
7$^*$ & 263.01389 & -34.75487 & 2.35 & -- & / & $^\dag2.50$ & NS & CCO & $0.95_{-0.01}^{+0.01}$ & 0.98 & $34.25^{+0.01}_{-0.01}$  \\
8 & 266.80853 & -30.03011 & 48.72 & / & / & / & / & / & $--_{---}^{+--}$ & -- & $--^{+--}_{---}$  \\
9 & 265.62632 & -28.74887 & 1.85 & 0.62 & 0.01 & 0.09 & Star & EclBin & $0.65_{-0.02}^{+0.01}$ & 1.42 & $30.16^{+0.01}_{-0.01}$  \\
10 & 265.14406 & -29.62898 & 1.63 & 0.46 & 0.01 & 0.18 & Star & YSO? & $1.35_{-0.19}^{+0.19}$ & 1.29 & $31.01^{+0.01}_{-0.02}$  \\
11 & 264.8929 & -29.1302 & 130.57 & / & / & / & / & / & $--_{---}^{+--}$ & -- & $--^{+--}_{---}$  \\
12 & 264.60944 & -29.03084 & 2.05 & 1.81 & 0.12 & 0.1 & Star & RotV* & $0.90_{-0.02}^{+0.03}$ & 1.14 & $30.09^{+0.01}_{-0.01}$  \\
13 & 266.47511 & -31.2509 & 0.91 & 0.32 & 0.01 & 0.38 & Star & Star & $1.19_{-0.04}^{+0.05}$ & 1.10 & $30.81^{+0.01}_{-0.02}$  \\
14 & 265.56239 & -29.24971 & 2.5 & 0.16 & 0.01 & 0.08 & Star & YSO? & $0.42_{-0.02}^{+0.02}$ & 1.06 & $29.27^{+0.06}_{-0.02}$  \\
15 & 264.12122 & -29.17493 & 2.81 & 0.68 & 0.03 & 0.71 & Star & Star & $1.86_{-0.32}^{+0.45}$ & 0.92 & $31.55^{+0.04}_{-0.04}$  \\
16 & 265.09948 & -28.94791 & 1.54 & 0.52 & 0.01 & 0.09 & Star & YSO? & $0.50_{-0.01}^{+0.02}$ & 0.94 & $29.67^{+0.02}_{-0.02}$  \\
17 & 264.83511 & -29.53097 & 1.81 & 0.23 & 0.00 & 0.13 & Star & Star & $0.60_{-0.02}^{+0.02}$ & 0.95 & $29.51^{+0.03}_{-0.03}$  \\
18 & 267.0848 & -30.47654 & 2.85 & 0.87 & 0.02 & 0.15 & Star & Star & $0.61_{-0.05}^{+0.04}$ & 1.09 & $29.62^{+0.08}_{-0.05}$  \\
19 & 265.42372 & -28.5567 & 6.04 & 1.88 & 0.07 & 0.11 & Star & HighPM* & $0.79_{-0.06}^{+0.05}$ & 1.03 & $29.37^{+0.04}_{-0.03}$  \\
20 & 265.49109 & -29.0927 & 1.65 & 0.64 & 0.01 & 0.13 & Star & Star & $0.59_{-0.01}^{+0.01}$ & 0.95 & $29.42^{+0.02}_{-0.02}$  \\
21 & 266.33712 & -29.24898 & 490.34 & / & / & / & / & / & $--_{---}^{+--}$ & -- & $--^{+--}_{---}$  \\
22 & 265.81399 & -31.26895 & 8.35 & 0.95 & 0.03 & 0.33 & Star & -- & $0.80_{-0.10}^{+0.08}$ & 0.90 & $30.04^{+0.09}_{-0.07}$  \\
23 & 265.85051 & -30.35429 & 4.53 & 2.9 & 0.05 & 0.4 & Star & -- & $0.71_{-0.14}^{+0.28}$ & 0.79 & $30.97^{+0.18}_{-0.13}$  \\
24 & 266.00836 & -31.1414 & 5.66 & 0.79 & 0.02 & 0.64 & Star & -- & $1.13_{-0.08}^{+0.10}$ & 0.86 & $30.92^{+0.02}_{-0.04}$  \\
25 & 266.84369 & -30.19861 & 49.43 & / & / & / & / & / & $--_{---}^{+--}$ & -- & $--^{+--}_{---}$  \\
26 & 265.06691 & -29.06044 & 3.34 & 0.7 & 0.00 & 1.3 & CV & CV & $> 57.9$ & 1.27 & $32.32^{+0.01}_{-0.01}$  \\
27 & 266.38902 & -30.98209 & 5.87 & 1.29 & 0.02 & 0.09 & Star & Star & $0.57_{-0.05}^{+0.05}$ & 1.11 & $29.34^{+0.03}_{-0.03}$  \\
28 & 266.75351 & -29.21974 & 5.31 & 0.29 & 0.00 & 1.29 & Star & -- & $1.11_{-0.07}^{+0.07}$ & 0.96 & $31.56^{+0.04}_{-0.04}$  \\
29 & 265.60418 & -29.94695 & 3.73 & 0.53 & 0.00 & 1.76 & Star & -- & $1.86_{-0.59}^{+1.23}$ & 0.70 & $31.15^{+0.05}_{-0.05}$  \\
30 & 265.57416 & -28.94655 & 6.44 & 0.28 & 0.00 & 5.24 & OBStar & Star & $0.95_{-0.05}^{+0.06}$ & 0.81 & $33.37^{+0.08}_{-0.08}$  \\
31 & 265.96371 & -28.77702 & 5.8 & 0.58 & 0.00 & 1 & Star & Star & $1.08_{-0.05}^{+0.06}$ & 0.90 & $31.91^{+0.03}_{-0.03}$  \\
32 & 264.82596 & -29.397 & 4.84 & -- & / & -- & LMXB? & XrayS & $^\ddag 1.55^{+0.17}_{-0.16}$ & 1.00 & $--^{+--}_{---}$  \\
33 & 266.556 & -29.4138 & 141.54 & / & / & / & / & / & $--_{---}^{+--}$ & -- & $--^{+--}_{---}$  \\
34 & 265.39028 & -28.67613 & 1.13 & 1.61 & 0.05 & 0.78 & CV? & XrayS & $3.99_{-0.36}^{+0.40}$ & 0.98 & $32.97^{+0.02}_{-0.02}$  \\
35 & 265.87392 & -30.13664 & 2.79 & 0.61 & 0.01 & 0.27 & Star & EclBin & $1.04_{-0.04}^{+0.04}$ & 0.8 & $29.92^{+0.03}_{-0.03}$  \\
36 & 265.83094 & -29.23332 & 8.38 & 0.14 & 0.00 & 3.44 & SySt? & XrayS & $1.12_{-0.09}^{+0.10}$ & 0.97 & $33.08^{+0.10}_{-0.08}$  \\
37 & 265.18975 & -29.27609 & 3.04 & 0.52 & 0.00 & 0.54 & CV? & XrayS & $5.19_{-0.83}^{+1.05}$ & 0.97 & $31.70^{+0.03}_{-0.03}$  \\
38 & 265.03794 & -28.79042 & 7.4 & 0.48 & 0.02 & 1.39 & CV & CV & $>41.59$ & 1.36 & $32.20^{+0.02}_{-0.02}$  \\
39 & 264.79975 & -31.19149 & 4.11 & 0.77 & 0.01 & 0.67 & Star & EclBin & $0.26_{-0.05}^{+0.82}$ & 0.92 & $30.33^{+0.09}_{-0.07}$  \\
40 & 264.22902 & -31.34948 & 66.89 & / & / & / & / & / & $--_{---}^{+--}$ & -- & $--^{+--}_{---}$  \\
41 & 264.99814 & -31.19027 & 7.74 & 0.12 & 0.00 & 0.63 & Star & -- & $0.48_{-0.11}^{+0.11}$ & 0.95 & $30.74^{+0.31}_{-0.14}$  \\
42 & 265.45677 & -30.43554 & 5.42 & 0.24 & 0.00 & 0.4 & Star & YSO? & $0.92_{-0.06}^{+0.04}$ & 0.93 & $30.31^{+0.07}_{-0.06}$  \\
43 & 267.14774 & -29.95801 & 8.91 & 0.99 & 0.03 & 1.64 & $\gamma~\rm CAS?$ & Be* & $4.99_{-1.20}^{+2.55}$ & 1.00 & $31.88^{+0.03}_{-0.03}$  \\
44 & 267.32238 & -30.59765 & 7.66 & 0.81 & 0.04 & 0.05 & Star & HighPM* & $0.29_{-0.03}^{+0.04}$ & 1.01 & $28.40^{+0.15}_{-0.09}$  \\
45 & 265.40313 & -29.42602 & 4.74 & 0.5 & 0.00 & 0.44 & Star & EclBin & $0.70_{-0.10}^{+0.21}$ & 0.82 & $30.55^{+0.20}_{-0.24}$  \\
46 & 267.40627 & -30.4145 & 1.62 & 0.38 & 0.01 & 1.53 & Star & -- & $0.92_{-0.50}^{+0.76}$ & 0.93 & $31.00^{+0.15}_{-0.08}$  \\
47 & 266.11016 & -29.05815 & 4.14 & 1.08 & 0.02 & 0.17 & Star & -- & $0.70_{-0.08}^{+0.06}$ & 0.90 & $29.18^{+0.07}_{-0.04}$  \\
48 & 266.42065 & -29.26281 & 653.95 & / & / & / & / & / & $--_{---}^{+--}$ & -- & $--^{+--}_{---}$  \\
49 & 267.58596 &  -30.4477 & 2.6 & 0.6 & 0.04 & 0.52 & Star & -- & $1.01_{-0.04}^{+0.05}$ & 1.12 & $31.00^{+0.02}_{-0.02}$  \\
50 & 267.94933 & -30.17975 & 8.6 & 1.25 & 0.23 & 0.18 & LP? & LP? & $0.89_{-0.04}^{+0.04}$ & 0.91 & $29.78^{+0.03}_{-0.02}$  \\
51 & 267.72457 & -30.38978 & 6.27 & 0.74 & 0.06 & 0.45 & Star & Star & $0.89_{-0.46}^{+0.20}$ & 1.00 & $30.44^{+0.06}_{-0.10}$  \\
52 & 267.48995 & -29.63616 & 60.85 & / & / & / & / & / & $--_{---}^{+--}$ & -- & $--^{+--}_{---}$  \\
53 & 268.0239 & -30.5066 & 2.42 & 1.09 & 0.20 & 1.79 & CV? & XrayS & $15.05_{-3.48}^{+33.96}$ & 0.79 & $32.09^{+0.01}_{-0.02}$  \\
54 & 267.64007 & -31.1554 & 382.78 & / & / & / & / & / & $--_{---}^{+--}$ & -- & $--^{+--}_{---}$  \\
55 & 261.12125 & -34.24733 & 6.25 & 1.89 & 0.01 & 2.03 & OBStar & YSO & $1.21_{-0.40}^{+0.68}$ & 0.97 & $31.53^{+3.19}_{-0.28}$  \\
56 & 261.13522 & -34.46266 & 2.24 & 1.15 & 0.00 & 0.14 & Star & Star & $0.52_{-0.19}^{+0.14}$ & 0.81 & $29.24^{+0.56}_{-0.12}$  \\
57 & 260.62038 & -37.44868 & 2.19 & 0.97 & 0.02 & 0.12 & Star & RotV* & $0.84_{-0.04}^{+0.03}$ & 1.17 & $29.99^{+0.02}_{-0.02}$  \\
58 & 261.67365 & -35.68165 & 2.56 & 1.62 & 0.01 & 0.14 & Star & RotV* & $0.84_{-0.02}^{+0.02}$ & 1.11 & $29.98^{+0.02}_{-0.02}$  \\
59 & 259.80462 & -36.85679 & 0.74 & 0.13 & 0.00 & 0.13 & Star & RotV* & $--_{---}^{+--}$ & -- & $--^{+--}_{---}$  \\
60 & 260.13187 & -35.85296 & 4.52 & 1.62 & 0.02 & 2.03 & OBStar & YSO & $2.22_{-0.17}^{+0.16}$ & 1.01 & $32.82^{+0.03}_{-0.02}$  \\
61 & 261.49385 & -35.54204 & 2.54 & 0.32 & 0.00 & 0.87 & CV & CV & $0.76_{-0.10}^{+0.15}$ & 0.97 & $31.07^{+0.13}_{-0.19}$  \\
62 & 260.52445 & -36.29416 & 5.21 & 2.02 & 0.03 & 0.14 & Star & -- & $0.84_{-0.12}^{+0.08}$ & 1.05 & $29.40^{+0.07}_{-0.04}$  \\
63 & 260.65281 & -36.34575 & 1.75 & 1.99 & 0.02 & 0.14 & Star & YSO? & $0.72_{-0.05}^{+0.06}$ & 0.96 & $29.30^{+0.04}_{-0.04}$  \\
64 & 260.2266 & -37.11441 & 229.66 & / & / & / & / & / & $--_{---}^{+--}$ & -- & $--^{+--}_{---}$  \\
65 & 260.64277 & -36.79915 & 1.39 & 1.71 & 0.02 & 0.84 & Star & -- & $1.79_{-0.12}^{+0.10}$ & 0.76 & $31.00^{+0.05}_{-0.04}$  \\
66 & 261.45934 & -35.51529 & 4.27 & 1.09 & 0.00 & 0.31 & Star & Star & $1.04_{-0.06}^{+0.06}$ & 1.02 & $30.29^{+0.05}_{-0.05}$  \\
67 & 259.81484 & -36.25924 & 4.39 & 1.85 & 0.01 & 0.14 & Star & YSO? & $0.69_{-0.12}^{+0.05}$ & 1.04 & $29.38^{+0.09}_{-0.03}$  \\
68 & 261.08508 & -35.35557 & 4.55 & 0.86 & 0.01 & 1.59 & AB? & -- & $7.16_{-3.90}^{+3.91}$ & 0.78 & $31.47^{+0.13}_{-0.11}$  \\
69 & 261.5216 & -36.74913 & 3.69 & 1.54 & 0.02 & 1.34 & Star & -- & $1.16_{-0.43}^{+0.04}$ & 0.97 & $30.97^{+0.09}_{-0.06}$  \\
70 & 259.80516 & -36.7682 & 7.92 & 2.19 & 0.03 & 0.13 & Star & YSO? & $1.18_{-0.24}^{+2.62}$ & 0.87 & $28.83^{+0.13}_{-0.08}$  \\
71 & 261.10059 & -36.60947 & 5.59 & 2.18 & 0.04 & 0.61 & Star & -- & $0.88_{-0.14}^{+0.11}$ & 1.01 & $30.55^{+0.14}_{-0.07}$  \\
72 & 260.1097 & -35.73556 & 3.2 & 2.11 & 0.04 & 1.55 & $\gamma~\rm CAS?$ & Be* & $6.27_{-1.63}^{+3.10}$ & 0.90 & $32.09^{+0.03}_{-0.03}$  \\
73 & 260.87917 & -34.74049 & 2.56 & 0.41 & 0.00 & 0.13 & Star & YSO? & $0.76_{-0.08}^{+0.07}$ & 1.19 & $29.04^{+0.06}_{-0.04}$  \\
74 & 260.85342 & -35.86781 & 3.44 & 0.87 & 0.00 & 1.31 & Star & -- & $0.96_{-0.27}^{+2.00}$ & 0.88 & $31.31^{+0.28}_{-0.15}$  \\
75 & 261.68921 & -35.03525 & 7.63 & 2.64 & 0.04 & 0.16 & Star & -- & $0.83_{-0.04}^{+0.05}$ & 0.95 & $29.26^{+0.05}_{-0.05}$  \\
76 & 261.58655 & -36.17202 & 11.09 & 1.26 & 0.01 & 0.98 & Star & -- & $1.40_{-0.31}^{+0.47}$ & 1.02 & $31.11^{+0.08}_{-0.06}$  \\
77 & 259.67512 & -36.793 & 154.68 & / & / & / & / & -- & $--_{---}^{+--}$ & -- & $--^{+--}_{---}$  \\
78 & 259.52853 & -36.18373 & 4.67 & -- & / & -- & Star & -- & $0.65_{-0.13}^{+0.13}$ & 0.78 & $--^{+--}_{---}$  \\
79 & 259.93532 & -35.89481 & 12.68 & 1.18 & 0.01 & 0.19 & Star & Star & $0.81_{-0.13}^{+0.11}$ & 1.15 & $29.07^{+0.09}_{-0.07}$  \\
80 & 260.56525 & -35.49181 & 2.28 & 1.59 & 0.02 & 1.21 & AB? & -- & $3.44_{-1.87}^{+3.60}$ & 1.07 & $30.94^{+0.11}_{-0.16}$  \\
81 & 260.91472 & -35.66941 & 1.99 & 0.56 & 0.00 & 1.2 & Star & -- & $1.12_{-0.88}^{+2.04}$ & 0.77 & $31.05^{+0.00}_{-0.21}$  \\
82 & 261.60676 & -35.27301 & 2.43 & 1.11 & 0.01 & 1.19 & Star & -- & $1.07_{-0.12}^{+0.14}$ & 0.88 & $31.29^{+0.12}_{-0.08}$  \\
83 & 260.85829 & -37.00325 & 42.06 & / & / & / & / & / & $--_{---}^{+--}$ & -- & $--^{+--}_{---}$  \\
84 & 259.81158 & -36.18805 & 4.05 & 2.24 & 0.03 & 0.19 & Star & -- & $0.80_{-0.05}^{+0.05}$ & 0.94 & $29.54^{+0.04}_{-0.04}$  \\
85 & 261.71776 & -35.96889 & 6.32 & 0.62 & 0.01 & 1.04 & Star & -- & $1.33_{-0.00}^{+0.00}$ & 0.94 & $30.34^{+0.00}_{-0.00}$  \\
86 & 260.66787 & -35.27488 & 2.02 & 0.31 & 0.00 & 0.16 & Star & -- & $1.06_{-0.11}^{+0.15}$ & 0.89 & $29.17^{+0.06}_{-0.06}$  \\
87 & 260.60084 & -34.84669 & 2.47 & 2.81 & 0.05 & 0.16 & Star & Star & $0.36_{-0.09}^{+0.15}$ & 1.16 & $28.91^{+0.37}_{-0.14}$  \\
88 & 261.34339 & -36.40743 & 8.64 & 0.79 & 0.01 & 1.26 & Star & -- & $1.50_{-0.13}^{+0.25}$ & 1.05 & $31.41^{+0.03}_{-0.03}$  \\
89 & 260.32565 & -36.1483 & 2.22 & 1.58 & 0.03 & 0.15 & Star & YSO? & $0.19_{-0.03}^{+0.12}$ & 0.92 & $30.31^{+0.74}_{-0.36}$  \\
90 & 263.83574 & -31.23724 & 5.32 & 0.54 & 0.02 & 0.38 & Star & Star & $0.86_{-0.05}^{+0.05}$ & 0.87 & $29.96^{+0.16}_{-0.11}$  \\
91 & 263.41858 & -31.37468 & 1.86 & 0.98 & 0.04 & 1.25 & Star & -- & $0.59_{-0.23}^{+4.83}$ & 0.82 & $31.99^{+0.81}_{-0.33}$  \\
92 & 263.67704 & -32.58167 & 0.7 & 0.05 & 0.00 & 0.86 & OBStar & Be* & $0.31_{-0.01}^{+0.02}$ & 1.02 & $32.40^{+0.02}_{-0.23}$  \\
93 & 262.63867 & -33.65427 & 0.79 & 1.11 & 0.05 & 0.36 & AB & RSCVnV* & $2.12_{-0.07}^{+0.18}$ & 1.25 & $31.25^{+0.01}_{-0.01}$  \\
94 & 261.8562 & -33.28015 & 3.82 & 0.74 & 0.04 & 0.12 & Star & RotV* & $0.89_{-0.03}^{+0.03}$ & 1.04 & $30.10^{+0.02}_{-0.02}$  \\
95 & 261.18254 & -34.19861 & 5.44 & 1.21 & 0.03 & -- & Star & YSO? & $1.30_{-0.14}^{+0.11}$ & 0.90 & $--^{+--}_{---}$  \\
96 & 263.05478 & -32.28329 & 0.59 & 1.01 & 0.01 & 0.1 & Star & Star & $0.58_{-0.02}^{+0.02}$ & 1.01 & $29.17^{+0.03}_{-0.01}$  \\
97 & 262.35647 & -33.7609 & 0.97 & 1.31 & 0.01 & 0.05 & Star & HighPM* & $0.48_{-0.03}^{+0.03}$ & 0.88 & $28.53^{+0.04}_{-0.03}$  \\
98 & 261.28752 & -34.1866 & 2.67 & 2.02 & 0.03 & 2.03 & OBStar & WolfRayet* & $1.48_{-0.15}^{+0.12}$ & 0.94 & $32.99^{+0.06}_{-0.04}$  \\
99 & 263.6595 & -33.47068 & 2.14 & 0.99 & 0.04 & 0.22 & Star & Star & $0.89_{-0.02}^{+0.02}$ & 1.13 & $30.45^{+0.02}_{-0.02}$  \\
100 & 263.16354 & -31.96606 & 4.76 & 0.63 & 0.00 & 0.27 & Star & -- & $0.99_{-0.03}^{+0.03}$ & 0.88 & $30.25^{+0.03}_{-0.03}$  \\
101 & 263.09815 & -34.33246 & 3.69 & 1.08 & 0.02 & 0.07 & Star & -- & $0.67_{-0.05}^{+0.05}$ & 0.77 & $28.80^{+0.05}_{-0.04}$  \\
102 & 263.89496 & -33.3681 & 0.78 & 1.56 & 0.02 & 0.29 & Star & Star & $--_{---}^{+--}$ & -- & $--^{+--}_{---}$  \\
103 & 264.36845 & -31.8273 & 4.46 & 0.27 & 0.00 & 0.22 & Star & Star & $0.70_{-0.04}^{+0.04}$ & 0.83 & $29.53^{+0.05}_{-0.04}$  \\
104 & 262.74025 & -33.51354 & 2.16 & 1.34 & 0.01 & 0.77 & AB? & -- & $3.04_{-0.21}^{+0.25}$ & 1.08 & $31.69^{+0.01}_{-0.01}$  \\
105 & 262.45071 & -32.56371 & 79.4 & / & / & / & / & / & $--_{---}^{+--}$ & -- & $--^{+--}_{---}$  \\
106 & 262.30385 & -31.53459 & 8.64 & 1.09 & 0.07 & 1.21 & OBStar & EclBin & $0.70_{-0.40}^{+0.60}$ & 1.03 & $30.68^{+0.24}_{-0.21}$  \\
107 & 261.3724 & -34.42094 & 3.83 & 2.61 & 0.04 & 1.81 & OBStar & YSO & $1.09_{-0.08}^{+0.10}$ & 1.10 & $30.71^{+0.59}_{-0.07}$  \\
108 & 263.03303 & -33.93703 & 173.74 & / & / & / & / & / & $--_{---}^{+--}$ & -- & $--^{+--}_{---}$  \\
109 & 263.93682 & -31.82402 & 4.64 & 0.18 & 0.00 & 0.14 & Star & Star & $0.71_{-0.03}^{+0.03}$ & 1.04 & $29.58^{+0.03}_{-0.03}$  \\
110 & 263.47878 & -33.32794 & 91.73 & / & / & / & / & -- & $--_{---}^{+--}$ & -- & $--^{+--}_{---}$  \\
111 & 263.47878 & -33.32794 & 35.6 & / & / & / & / & -- & $--_{---}^{+--}$ & -- & $--^{+--}_{---}$  \\
112 & 262.83628 & -33.8481 & 66.33 & / & / & / & / & / & $--_{---}^{+--}$ & -- & $--^{+--}_{---}$  \\
113 & 263.57693 & -34.11428 & 7.03 & 0.92 & 0.02 & 0.26 & Star & EclBin & $1.00_{-0.10}^{+0.10}$ & 0.72 & $30.15^{+0.06}_{-0.05}$  \\
114 & 264.51089 & -31.84742 & 2.12 & 0.25 & 0.00 & 0.42 & Star & Star & $0.83_{-0.04}^{+0.04}$ & 0.83 & $30.29^{+0.05}_{-0.05}$  \\
115 & 263.68402 & -32.61845 & 29.41 & / & / & / & / & / & $--_{---}^{+--}$ & -- & $--^{+--}_{---}$  \\
116 & 262.43828 & -33.84158 & 3.12 & 1.23 & 0.02 & 0.83 & SySt? & Star & $1.64_{-0.12}^{+0.10}$ & 0.90 & $32.04^{+0.07}_{-0.04}$  \\
117 & 263.69848 & -32.68514 & 11.48 & 0.18 & 0.00 & 0.32 & Star & Star & $0.83_{-0.23}^{+0.38}$ & 1.20 & $29.48^{+0.82}_{-0.07}$  \\
118 & 264.58478 & -32.89498 & 2.23 & 1.2 & 0.01 & 0.66 & Star & Star & $0.90_{-0.06}^{+0.06}$ & 0.90 & $30.71^{+0.30}_{-0.05}$  \\
119 & 263.32608 & -33.25912 & 176.32 & / & / & / & / & / & $--_{---}^{+--}$ & -- & $--^{+--}_{---}$  \\
120 & 263.08604 & -33.33976 & 5.66 & 1.68 & 0.03 & 1.67 & OBStar & Star & $0.83_{-0.10}^{+0.08}$ & 0.83 & $31.97^{+0.10}_{-0.07}$  \\
121 & 263.34932 & -33.90866 & 14.72 & 1.9 & 0.04 & 0.65 & Star & -- & $0.83_{-0.15}^{+0.10}$ & 0.81 & $30.69^{+0.15}_{-0.08}$  \\
122 & 263.70078 & -32.53162 & 13.19 & 2.03 & 0.03 & 1.18 & Star & Star & $1.00_{-0.13}^{+0.43}$ & 1.26 & $31.28^{+0.31}_{-0.12}$  \\
123 & 262.80793 & -32.36837 & 4.73 & 2.7 & 0.17 & 3.4 & CV? & XrayS & $64.0_{-11.2}^{+0}$ & 1.03 & $32.98^{+0.02}_{-0.02}$  \\
124 & 264.43788 & -32.58557 & 3.78 & 0.22 & 0 & 0.45 & Star & Star & $0.90_{-0.06}^{+0.06}$ & 0.90 & $30.54^{+0.07}_{-0.06}$  \\
125 & 264.34576 & -33.0397 & 3.14 & 0.78 & 0.03 & 0.22 & Star & -- & $0.66_{-0.14}^{+0.18}$ & 1.00 & $29.42^{+0.27}_{-0.07}$  \\
126 & 262.65302 & -33.53992 & 8.07 & 1.29 & 0.02 & 0.69 & Star & -- & $0.89_{-0.10}^{+0.11}$ & 1.06 & $30.83^{+0.09}_{-0.07}$  \\
127 & 264.55697 & -31.91006 & 4.78 & 0.23 & 0.00 & 0.41 & Star & -- & $0.89_{-0.08}^{+0.04}$ & 1.00 & $30.52^{+5.39}_{-0.09}$  \\
128 & 264.59869 & -32.00864 & 7.93 & 0.45 & 0.01 & 0.06 & Star & Star & $0.51_{-0.04}^{+0.04}$ & 1.03 & $28.44^{+0.02}_{-0.02}$  \\
129 & 262.46605 & -32.63936 & 7.88 & 2.95 & 0.08 & -- & Star & -- & $0.15_{-0.09}^{+0.16}$ & 1.36 & $--^{+--}_{---}$  \\
130 & 263.77262 & -32.57439 & 7.6 & 1.52 & 0.03 & 1.68 & Star & Star & $0.99_{-0.6}^{+3.49}$ & 1.26 & $30.22^{+0.21}_{-0.11}$  \\
131 & 264.06367 & -32.19279 & 1.35 & 1.04 & 0.01 & 1.00 & Star & -- & $0.67_{-0.43}^{+0.27}$ & 0.58 & $30.20^{+3.37}_{-0.20}$  \\
132 & 261.61041 & -34.28325 & 10.72 & 1.42 & 0.01 & 1.6 & OBStar & Star & $0.85_{-0.10}^{+0.09}$ & 0.73 & $31.73^{+0.18}_{-0.15}$  \\
133 & 263.69942 & -32.57281 & 3.2 & 1.22 & 0.02 & 1.09 & Star & Star & $--_{---}^{+--}$ & -- & $--^{+--}_{---}$  \\
134 & 263.7094 & -32.65954 & 13.17 & 0.96 & 0.01 & 1.41 & Star & Star & $0.10_{-0.04}^{+0.13}$ & 1.12 & $31.53^{+0.53}_{-0.36}$  \\
135 & 264.56809 & -32.37547 & 5.01 & 0.4 & 0.01 & 0.46 & Star & Star & $0.94_{-0.07}^{+0.06}$ & 1.20 & $30.44^{+0.07}_{-0.05}$  \\
136 & 264.36713 & -31.51569 & 2.35 & 0.27 & 0.00 & 1.41 & LP? & LP? & $3.39_{-0.30}^{+0.35}$ & 0.93 & $32.35^{+0.02}_{-0.02}$  \\
137 & 263.74624 & -33.61479 & 14.29 & 0.34 & 0.02 & 0.24 & Star & Star & $0.46_{-0.17}^{+0.25}$ & 0.92 & $29.25^{+0.19}_{-0.16}$  \\
138 & 263.55815 & -31.74728 & 3.94 & 0.43 & 0.00 & -- & CV? & XrayS & $9.66_{-3.16}^{+5.47}$ & 1.23 & $--^{+--}_{---}$  \\
139 & 263.14006 & -32.9542 & 9.81 & 1.44 & 0.02 & 1.12 & Star & -- & $0.73_{-0.63}^{+0.26}$ & 0.72 & $31.63^{+0.51}_{-0.21}$  \\
140 & 261.21152 & -34.39249 & 49.22 & / & / & / & / & / & $--_{---}^{+--}$ & -- & $--^{+--}_{---}$  \\
141 & 264.64286 & -32.17214 & 4.85 & 2.58 & 0.06 & -- & Star & -- & $1.12_{-0.21}^{+0.29}$ & 0.64 & $--^{+--}_{---}$  \\
142 & 263.14973 & -33.22873 & 2.65 & 0.8 & 0.00 & 0.38 & Star & YSO? & $0.92_{-0.20}^{+0.41}$ & 0.78 & $29.92^{+0.29}_{-0.17}$  \\
143 & 261.32923 & -34.2492 & 16.21 & 2.76 & 0.04 & 0.11 & Star & Star & $0.25_{-0.13}^{+0.20}$ & 0.55 & $30.63^{+0.85}_{-0.31}$  \\
144 & 263.58108 & -34.14612 & 11.07 & -- & / & -- & CV? & -- & $5.88_{-2.37}^{+6.02}$ & 0.64 & $--^{+--}_{---}$  \\
145 & 263.50989 & -33.35568 & 60.17 & / & / & / & / & / & $--_{---}^{+--}$ & -- & $--^{+--}_{---}$  \\
146 & 263.94281 & -32.11927 & 3.42 & 0.49 & 0.01 & 1.15 & CV? & -- & $4.07_{-0.69}^{+0.77}$ & 0.93 & $31.76^{+0.03}_{-0.03}$  \\
147 & 262.84051 & -32.57101 & 10.82 & -- & / & -- & Star & -- & $1.25_{-0.29}^{+0.57}$ & 0.73 & $--^{+--}_{---}$  \\
148 & 263.67559 & -32.56472 & 31.59 & / & / & / & / & / & $--_{---}^{+--}$ & -- & $--^{+--}_{---}$  \\
149 & 261.52052 & -34.29521 & 55.35 & / & / & / & / & / & $--_{---}^{+--}$ & -- & $--^{+--}_{---}$  \\
150 & 262.66742 & -33.93408 & 73.07 & / & / & / & / & / & $--_{---}^{+--}$ & -- & $--^{+--}_{---}$  \\
151 & 262.89866 & -33.32885 & 1.96 & 1.68 & 0.02 & 0.64 & Star & -- & $0.96_{-0.35}^{+0.25}$ & 1.07 & $30.25^{+0.46}_{-0.16}$  \\
152 & 264.39011 & -33.54535 & 6.81 & 0.58 & 0.01 & 0.9 & LP? & LP? & $0.80_{-0.15}^{+0.16}$ & 0.74 & $30.90^{+0.16}_{-0.10}$  \\
153 & 262.62761 & -34.53895 & 1.54 & 0.54 & 0.00 & 0.39 & AB & RSCVnV* & $2.60_{-0.05}^{+0.05}$ & 1.45 & $31.78^{+0.00}_{-0.00}$  \\
154 & 261.9983 & -36.12627 & 0.17 & 0.08 & 0.00 & 0.13 & Star & YSO? & $0.42_{-0.04}^{+0.08}$ & 1.04 & $29.94^{+0.01}_{-0.01}$  \\
155 & 262.24898 & -35.27784 & 1.27 & 1.91 & 0.02 & 0.11 & Star & -- & $0.63_{-0.07}^{+0.07}$ & 1.22 & $29.74^{+0.05}_{-0.04}$  \\
156 & 263.29425 & -34.61242 & 123.67 & / & / & / & / & / & $--_{---}^{+--}$ & -- & $--^{+--}_{---}$  \\
157 & 263.12137 & -34.65918 & 268.41 & / & / & / & / & / & $--_{---}^{+--}$ & -- & $--^{+--}_{---}$  \\
158 & 261.9384 & -35.83948 & 1.92 & 0.48 & 0.00 & 1.76 & OBStar & Star & $1.00_{-0.17}^{+0.31}$ & 1.18 & $31.47^{+0.16}_{-0.29}$  \\
159 & 262.83494 & -35.14534 & 2.01 & 0.48 & 0.01 & 0.49 & Star & Star & $0.93_{-0.14}^{+0.06}$ & 1.07 & $30.12^{+1.70}_{-0.04}$  \\
160 & 263.15593 & -34.79799 & 109.64 & / & / & / & / & / & $--_{---}^{+--}$ & -- & $--^{+--}_{---}$  \\
161 & 262.66104 & -34.85446 & 11.34 & 0.13 & 0.00 & 0.51 & Star & -- & $0.70_{-0.25}^{+0.11}$ & 0.75 & $29.78^{+3.63}_{-0.18}$  \\
162 & 262.05925 & -34.69005 & 3.39 & 1.28 & 0.01 & 0.24 & Star & Star & $0.39_{-0.12}^{+0.23}$ & 1.19 & $29.49^{+2.68}_{-0.27}$  \\
163 & 262.62358 & -35.10835 & 3.58 & 1.81 & 0.01 & 0.79 & Star & -- & $0.43_{-0.27}^{+4.05}$ & 0.79 & $30.57^{+0.90}_{-0.26}$  \\
164 & 263.08261 & -34.84451 & 51.9 & / & / & / & / & / & $--_{---}^{+--}$ & -- & $--^{+--}_{---}$  \\
165 & 262.7455 & -35.13685 & 7.03 & 1.66 & 0.02 & -- & CV & CV & $16.01_{-3.35}^{+3.36}$ & 0.97 & $--^{+--}_{---}$  \\
166 & 266.67037 & -32.23414 & 5.98 & 3.33 & 0.07 & 0.01 & Star & HighPM* & $0.52_{-0.12}^{+0.07}$ & 0.93 & $27.03^{+0.14}_{-0.05}$  \\
167 & 266.59903 & -31.58362 & 2.37 & 1.18 & 0.02 & 0.13 & Star & YSO? & $0.81_{-0.04}^{+0.04}$ & 0.96 & $29.58^{+0.02}_{-0.02}$  \\
168 & 266.55923 & -32.10299 & 8.65 & 3.53 & 0.10 & 0.01 & Star & HighPM* & $0.27_{-0.02}^{+0.02}$ & 1.04 & $28.33^{+0.04}_{-0.03}$  \\
169 & 264.91851 & -32.23852 & 1.53 & 0.11 & 0.00 & 0.46 & Star & Star & $0.87_{-0.04}^{+0.04}$ & 0.85 & $30.75^{+0.04}_{-0.03}$  \\
170 & 266.25999 & -31.9929 & 7.4 & -- & / & -- & Star & Star & $1.04_{-0.13}^{+0.13}$ & 0.95 & $--^{+--}_{---}$  \\
171 & 265.01482 & -32.39085 & 3.05 & 0.74 & 0.01 & 0.46 & Star & Star & $0.82_{-0.05}^{+0.04}$ & 0.74 & $30.60^{+0.05}_{-0.04}$  \\
172 & 264.80366 & -31.78112 & 5.07 & 0.35 & 0.00 & 0.47 & Star & Star & $0.80_{-0.04}^{+0.05}$ & 0.91 & $30.13^{+0.10}_{-0.10}$  \\
173 & 265.05627 & -32.28702 & 9.21 & 2.64 & 0.16 & 1.79 & CV? & -- & $4.05_{-1.46}^{+7.58}$ & 1.35 & $30.95^{+0.15}_{-0.10}$  \\
174 & 265.05089 & -32.26833 & 2.51 & 0.48 & 0.00 & 0.47 & Star & Star & $0.84_{-0.05}^{+0.05}$ & 1.03 & $30.01^{+0.15}_{-0.07}$  \\
175 & 265.13513 & -32.33585 & 5.47 & 1.37 & 0.02 & 0.47 & Star & Star & $1.12_{-0.10}^{+0.11}$ & 1.12 & $29.53^{+0.07}_{-0.06}$  \\
176 & 266.29555 & -32.44883 & 2.1 & 0.25 & 0.00 & 0.94 & Star & Star & $1.11_{-0.13}^{+0.15}$ & 0.71 & $30.86^{+0.12}_{-0.09}$  \\
177 & 265.17495 & -32.15717 & 3.79 & 1.00 & 0.01 & 0.45 & Star & Star & $1.08_{-0.46}^{+1.04}$ & 0.84 & $29.73^{+0.15}_{-0.11}$  \\
178 & 265.22077 & -32.04849 & 8.1 & 2.2 & 0.07 & 0.24 & Star & -- & $0.65_{-0.40}^{+0.24}$ & 1.28 & $29.33^{+0.39}_{-0.24}$  \\
179 & 265.14562 & -32.36907 & 6.07 & 0.55 & 0.00 & 0.45 & Star & -- & $0.18_{-0.23}^{+0.35}$ & 0.73 & $30.53^{+4.42}_{-0.43}$  \\
180 & 264.81231 & -32.40529 & 5.99 & 3.12 & 0.07 & 0.49 & Star & Star & $1.34_{-1.09}^{+1.09}$ & 0.94 & $29.79^{+0.00}_{-0.22}$  \\
181 & 264.89777 & -32.20737 & 7.36 & 0.42 & 0.00 & 0.54 & Star & Star & $0.85_{-0.04}^{+0.04}$ & 0.93 & $30.31^{+0.06}_{-0.05}$  \\
182 & 265.09786 & -32.35375 & 5.69 & 0.73 & 0.01 & 0.46 & Star & Star & $0.71_{-0.27}^{+0.09}$ & 0.70 & $30.37^{+0.50}_{-0.07}$  \\ 
\\
\hline
H1 & 265.97825	& -29.74572 & 1.41 & -- & -- & $^\dag$8.20 & HMXB & HMXB & $^\ddag 1.40^{+0.05}_{-0.05}$ & 0.94 & $36.91^{+0.02}_{-0.02}$  \\
H2 &266.81511 &	-29.96639 & 2.70 & -- & -- & $^\dag$2.50 & PSR & PSR & $^\ddag 1.77^{+0.05}_{-0.05}$ & 1.05 & $33.93^{+0.02}_{-0.02}$  \\
H3 & 260.28011 & -37.45432 & 95.33 & / & / & / & / & / & / & / & / \\

H4 & 266.39901 & -29.02644 & 783.87 & / & / & / & / & / & / & / & /  \\
H5 & 266.18910 & -29.84579 & 0.26 & -- & -- & $^\dag$8.20 & LMXB? & XrayS & $^\ddag 1.43^{+0.17}_{-0.16}$ & 0.91 & $34.76^{+0.07}_{-0.06}$  \\
H6 & 267.69408 & -31.27340 & 390.72 & / & / & / & / & / & / & / & /  \\
H7 & 263.86438 & -32.93106 & 2.86 & 0.73 & 0.03 & $^\dag$8.20 & HMXB & HMXB & $^\ddag 1.14^{+0.05}_{-0.05}$ & 1.22 & $35.28^{+0.01}_{-0.01}$

\end{longtable}
\label{tab:specsrc}
\tablefoot{
(1): Source identifier; Src 1-7 are labeled with $^*$ meaning they have ``strong'' association in the eRASS1 Hard catalogue.
H1-H7 represents the ``Hard-only'' sources;
(2)-(3): XMM‑Newton right ascension and declination (J2000);
(4): Angular separation between XMM and eROSITA positions (arcseconds);
(5): Angular separation between XMM and Gaia positions (arcseconds);
(6): The false-matched rate given the local Gaia source density, details in Sect.~\ref{sec:gaia};
(7): Photometric distance from Gaia DR3 (kpc); Sources labeled with $^\dag$ mean their distance are taken from literature.
(8): Source classification from this work (see Sect.~\ref{sec:class});
(9): Source type as listed in the SIMBAD (if available);
(10)–(12): Best-fit spectral parameters:mean plasma temperature $\rm kT_{\rm mean}$ (keV), reduced $\chi^2$, and unabsorbed 0.2--10 keV luminosity ($\rm erg~s^{-1}$).
Sources labeled with $^\ddag$ are those fitted by power-law model and the best-fitted $\Gamma$ are provided here.
}

\section{X-ray spectra plots for selected exotic sources}
\label{app:spec}

\begin{figure}
  \centering
   \includegraphics[width=1.03\hsize]{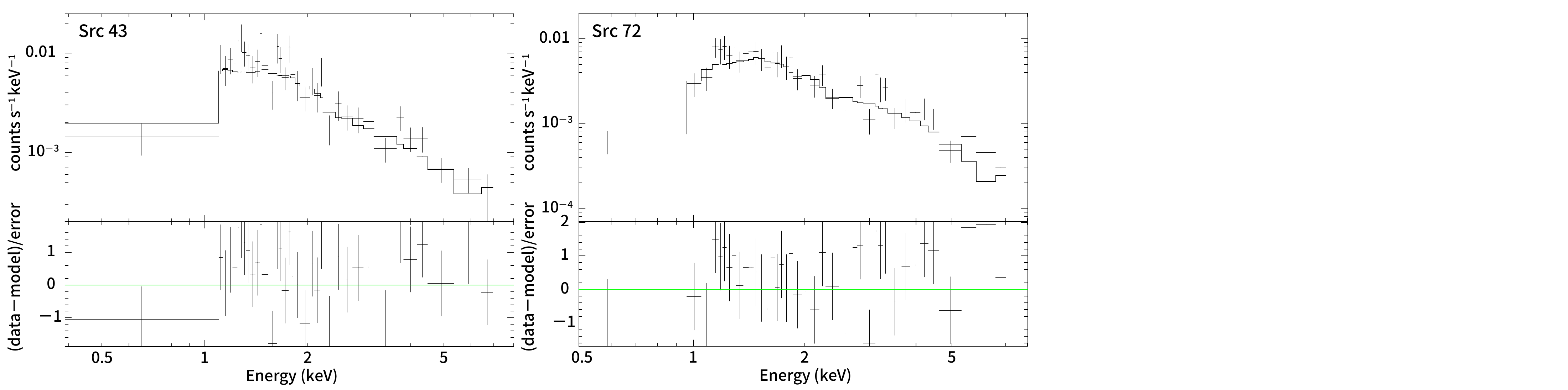}
      \caption[width=\hsize]{The spectra with best-fitted model for the two $\gamma$ CAS candidates.}
       \label{fig:yCASspec}
\end{figure}

\begin{figure}
  \centering
   \includegraphics[width=1.03\hsize]{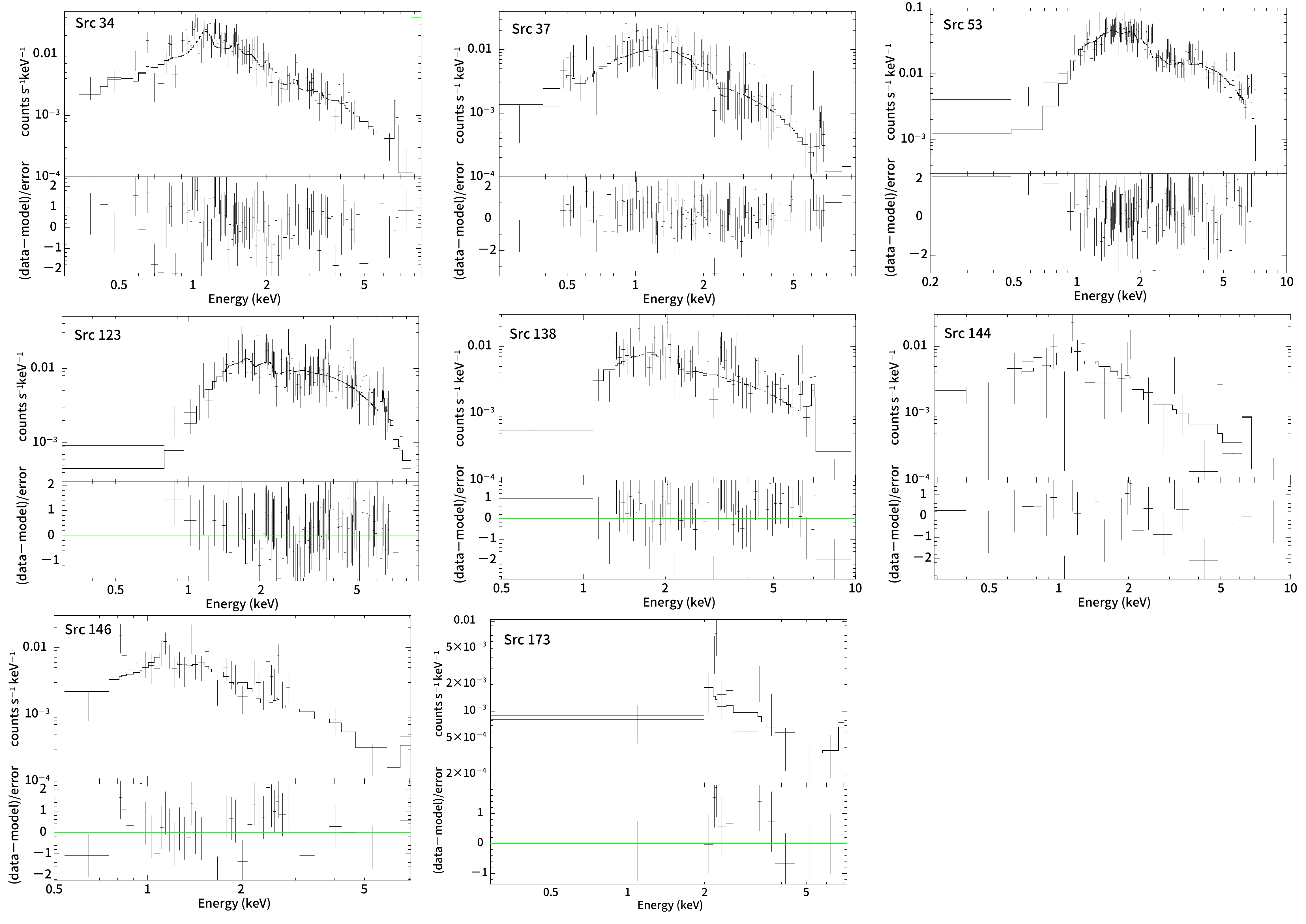}
      \caption[width=\hsize]{The spectra with best-fitted model for the eight CV candidates.}
       \label{fig:cvspec}
\end{figure}

\begin{figure}
  \centering
   \includegraphics[width=1.03\hsize]{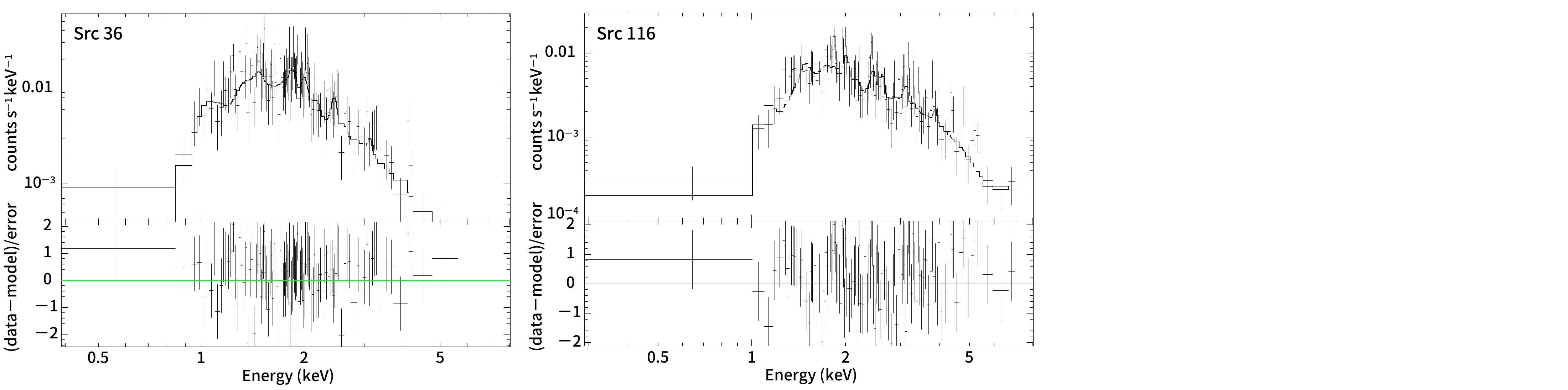}
      \caption[width=\hsize]{The spectra with best-fitted model for the two symbiotic star candidates.}
       \label{fig:systspec}
\end{figure}

\begin{figure}
  \centering
   \includegraphics[width=1.03\hsize]{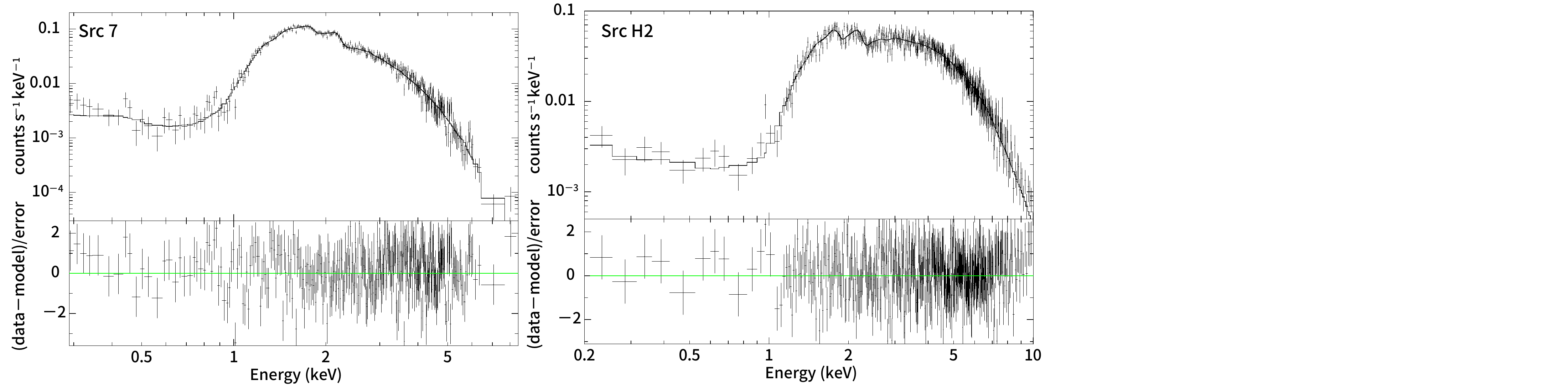}
      \caption[width=\hsize]{The spectra with best-fitted model for the two isolated NSs.}
       \label{fig:lmxbspec}
\end{figure}

\begin{figure}
  \centering
   \includegraphics[width=1.03\hsize]{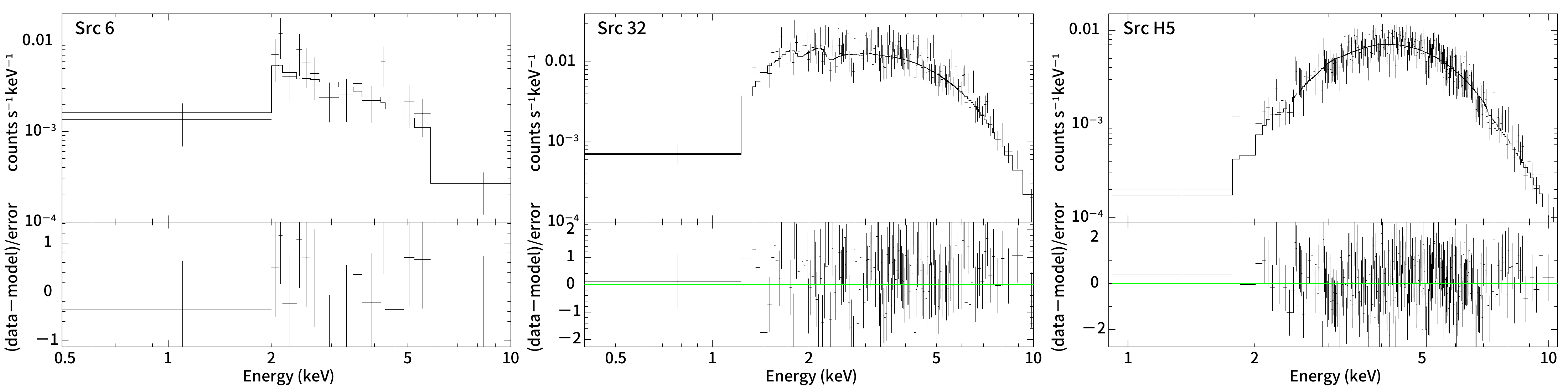}
      \caption[width=\hsize]{The spectra with best-fitted model for the three LMXB candidates.}
       \label{fig:nsspec}
\end{figure}

\begin{figure}
  \centering
   \includegraphics[width=1.03\hsize]{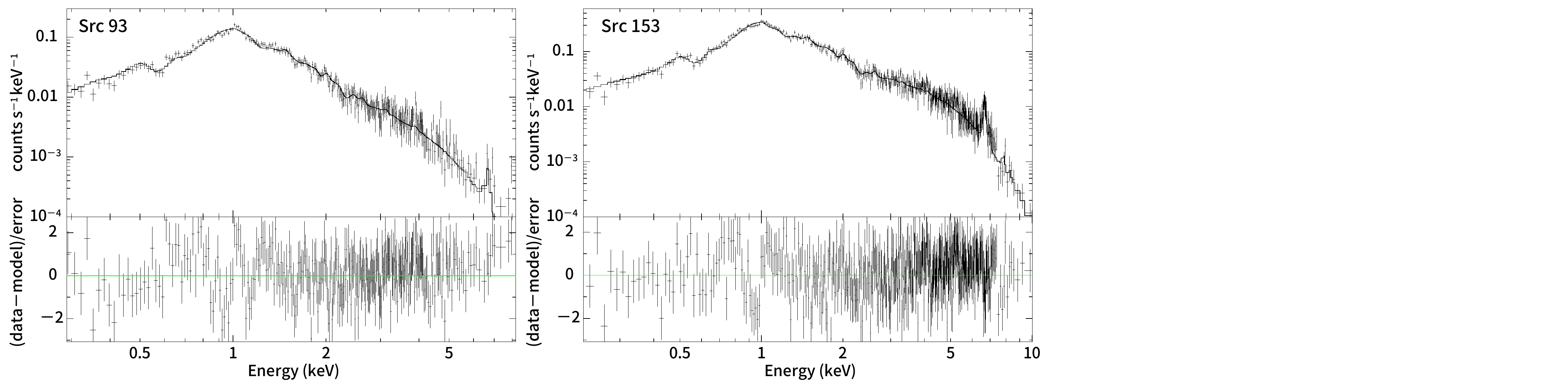}
      \caption[width=\hsize]{The spectra with best-fitted model for the two RS CVn systems.}
       \label{fig:RSCVnspec}
\end{figure}

\end{appendix}
\end{document}